\documentclass[10pt]{article}
\usepackage{a4wide}
\usepackage{color}
\usepackage{cite}
\usepackage{graphicx}

\usepackage{amssymb}
\usepackage{amsmath}
\usepackage{epstopdf}
\usepackage{enumitem}
\usepackage{multirow}

\newcommand{\dd}{{\rm{d}}}
\newcommand{\e}{\mathrm{e}}
\newcommand{\complex}{\mathrm{i}}

\def \bolde {\mbox{\boldmath$e$}}
\def \boldE {\mbox{\boldmath$E$}}
\def \boldk {\mbox{\boldmath$k$}}
\def \boldl {\mbox{\boldmath$l$}}
\def \boldm {\mbox{\boldmath$m$}}
\def \boldZ {\mbox{\boldmath$Z$}}
\def \boldU {\mbox{\boldmath$U$}}
\def \boldV {\mbox{\boldmath$V$}}
\def \boldW {\mbox{\boldmath$W$}}

\def \bF {\mathbf{F}}
\def \bA {\mathbf{A}}

\DeclareMathOperator{\sech}{sech}

\begin{document}

\title{Algebraic classification of 2+1 geometries: a new approach}

\author{
Mat\'u\v{s} Papaj\v{c}\'{\i}k
and
Ji\v{r}\'{\i} Podolsk\'y\thanks{{\tt
matus.papajcik@matfyz.cuni.cz,
podolsky@mbox.troja.mff.cuni.cz
}}
\\ \ \\ \ \\
Charles University, Faculty of Mathematics and Physics, \\
Institute of Theoretical Physics, \\
V~Hole\v{s}ovi\v{c}k\'ach 2, 18000 Prague 8, Czechia
}

\maketitle

\begin{abstract}
We present a convenient method of algebraic classification of 2+1 spacetimes into the types I, II, D, III, N and O, without using any field equations. It is based on the 2+1 analogue of the Newman--Penrose curvature scalars~$\Psi_{\rm A}$ of distinct boost weights, which are specific projections of the Cotton tensor onto a suitable null triad. The algebraic types are then simply determined by the gradual vanishing of such Cotton scalars, starting with those of the highest boost weight. This classification is directly related to the specific multiplicity of the Cotton-aligned null directions (CANDs) and to the corresponding Bel--Debever criteria. Using a bivector (that is 2-form) decomposition, we demonstrate that our method is fully equivalent to the usual Petrov-type classification of 2+1 spacetimes based on the eigenvalue problem and determining the respective canonical Jordan form of the Cotton--York tensor. We also derive a simple synoptic algorithm of algebraic classification based on the key polynomial curvature invariants. To show the practical usefulness of our approach, we perform the classification of several explicit examples, namely the general class of Robinson--Trautman spacetimes with an aligned electromagnetic field and a cosmological constant, and other metrics of various algebraic types.
\end{abstract}


\vfil\eject

\tableofcontents

\newpage


\section{Introduction}
\label{sec:Introduction}

Algebraic classification of spacetimes is an important tool for investigation and understanding of exact solutions of Einstein's field equations and other theories of gravity. In the context of ${D=4}$ \emph{general relativity} (that is for ${3+1}$ geometries) this was developed at the end of the 1950s by Petrov, G\'eh\'eniau, Pirani, Bell, Debever and Penrose \cite{Petrov:1954, Geheniau:1957, Pirani:1957, Bel:1959, Debever:1959a, Debever:1959b, Penrose:1960} using various equivalent approaches. In its most convenient formulation, related to the study of gravitational radiation (spacetimes of type~N) and also stationary black holes (of type~D), this is based on finding the multiplicity of four possible principal null directions (PNDs) of the Weyl curvature tensor, encoded in its null-frame components which are denoted as the complex Newman--Penrose scalars $\Psi_{\rm A}$, where ${{\rm A} = 0,1,2,3,4}$, see \cite{NewmanPenrose:1962}. Comprehensive reviews of this topic can be found in the monographs \cite{PenroseRindler:1984, Stephani}.

In 2004, this key concept of algebraic classification was extended to \emph{higher dimensions} ${D>4}$ by Coley, Milson, Pravda and Pravdov\'a \cite{ColeyMilsovPravdaPravdova:2004, MilsonColeyPravdaPravdova:2005}. In such a case, there are many more components of the Weyl tensor, but all their null-frame projections can again be sorted into just five groups with distinct boost weights. This fact enables one to perform the classification of the Weyl tensor in an analogous way as in the ${D=4}$ case, i.e., by the multiplicity of four Weyl-aligned null directions (WANDs), see the reviews \cite{Coley:2008, OrtaggioPravdaPravdova:2013}. To keep the closest possible analogy with the standard Newman--Penrose formalism, Krtou\v{s} and Podolsk\'y \cite{KrtousPodolsky:2006} introduced the familiar notation $\Psi_{\rm A}$ to represent all the relevant \emph{real} Weyl scalars in any ${D>4}$.

In fact, it should be emphasized that the classification scheme developed in \cite{MilsonColeyPravdaPravdova:2005} applies to \emph{any tensor} in \emph{arbitrary Lorentzian geometry}. Although not explicitly mentioned in this seminal work, it can be immediately observed that the scheme is valid also in the lower-dimensional case ${D=3}$ admitting two independent null directions and just one additional spatial direction.

From this general point of view, our classification method is an application of the scheme presented in \cite{MilsonColeyPravdaPravdova:2005, OrtaggioPravdaPravdova:2013} to 2+1 Lorentzian geometries in which we take the rank-3 \emph{Cotton tensor}  \cite{Cotton:1899} (instead of the \emph{identically vanishing} rank-4 Weyl tensor) as the key geometric quantity. The Cotton algebraic types correspond to the general classification into (primary) \emph{principal} and \emph{secondary alignment types} (PAT and SAT), as introduced for an arbitrary tensor by definitions 4.1 and 4.2 in \cite{MilsonColeyPravdaPravdova:2005}, and 2.5 in \cite{OrtaggioPravdaPravdova:2013}).

Classification of spacetimes in lower dimension ${D=3}$ was introduced many years ago. Neither the Petrov approach (based on the eigenvalue problem of the Weyl tensor) nor the Debever--Penrose analysis (based on the multiplicity of the Weyl tensor PNDs) could be directly applied because in 2+1 geometries the rank-4 Weyl tensor vanishes. Instead, it was found that the fundamental role for the algebraic classification plays the rank-3 Cotton tensor. The number of its independent components in 2+1 gravity is \emph{five}, so that it can be mapped onto the rank-2 symmetric and traceless \emph{Cotton--York tensor}. This tensor can be represented by a ${3 \times 3}$ matrix, and thus its algebraic classification can be performed analogously to the original Petrov approach. This was done in 1986 by Barrow, Burd and Lancaster \cite{Barrow}.

Such a classification in 2+1 gravity is, nevertheless, different from its ${D=4}$ counterpart. In the \emph{actual} formulation of the eigenvalue problem, the symmetry of the Cotton--York tensor is no longer manifest. The eigenvalues and also the corresponding eigenvectors can thus generally be complex. This feature was pointed out and remedied by Garc\'ia, Hehl, Heinicke and Mac\'ias in 2004. In their paper \cite{GHHM}, it was proposed to classify the spacetimes according to the possible \emph{Jordan forms} of the Cotton--York tensor in a suitable orthonormal basis. By this method, the spacetimes were divided into the types I, II, D, III, N and O. To deal with the possible complex eigenvalues, an additional type I$^\prime$ was proposed which restricts the solutions to only real numbers.

Alternative approaches to classification of 2+1 spacetimes were also presented. The formalism of \emph{null basis} was developed in \cite{Hall&Capocci}, while in \cite{Castillo&Ceballos} a \emph{spinor algebra} was established and used for the Ricci and Cotton--York tensors. An invariant \emph{Karlhede classification} method was developed in \cite{Sousa:2008} employing the Ricci and Cotton--York real spinors. Interestingly, in \emph{topologically massive gravity} (TMG), whose action involves a gravitational Chern--Simons term, the field equations imply that the Cotton--York tensor is proportional to the traceless Ricci tensor. Therefore, the Petrov-type classification of 2+1 spacetimes in TMG is equivalent to the Segre classification of the simpler traceless Ricci tensor, see \cite{ChowPopeSezgin:2010a, ChowPopeSezgin:2010b}.

Actually, the Segre--Pirani--Pleba\'nski classification of the \emph{energy-momentum tensor} of matter, related to the traceless Ricci tensor, is another important way of characterizing the spacetime. It takes advantage of its symmetry property, so that the eigenvalues and eigenvectors can be directly determined by a standard procedure, and classified using the nomenclature of Pleba\'nski \cite{Plebanski:1964}. More details on these schemes, and their application to many important classes of exact solutions to 2+1 gravity, are given in the monograph \cite{Garcia:2017}, see in particular sections 1.2 and 20.5 therein.

In our work, we now propose a \emph{simpler and general method} of algebraic classification of spacetimes in 2+1 gravity which does not assume any field equations. It is based \emph{directly} on the Cotton tensor, namely on \emph{five Cotton scalars}~$\Psi_{\rm A}$ obtained by specific projections onto a null triad. In fact, this is a lower-dimensional analogue of the standard Newman--Penrose method of ${D=4}$ general relativity which uses the Weyl tensor. It is naturally related to the multiplicity of the Cotton-aligned null directions (CANDs), in full analogy to the multiplicity of PNDs and WANDs. We show that this approach is equivalent to the classification developed in \cite{GHHM} which relies on the canonical Jordan forms of the Cotton--York tensor. We also identify key scalar polynomial invariants constructed from the Cotton scalars~$\Psi_{\rm A}$, which conveniently assist with the algebraic classification.

We begin in Sec.~\ref{sec:Cotton} by establishing the notation and introducing the Cotton tensor~$C_{abc}$. In subsequent Sec.~\ref{sec:PsiAdef} we define a null triad onto which the Cotton tensor is projected, obtaining thus the key Newman--Penrose-type Cotton scalars~$\Psi_{\rm A}$. This allows us to present a very simple classification scheme in Sec.~\ref{sec:PsiAclassification}. Then in Sec.~\ref{sec:Cottonbivector} we define a bivector basis and prove that the corresponding components of the Cotton tensor are just the scalars~$\Psi_{\rm A}$. Relation to the Bel--Debever criteria for the privileged aligned null vector $\boldk$ is demonstrated in Sec.~\ref{Bel-Debever-criteria}.
All Lorentz transformations are investigated in Sec.~\ref{Lorentz-transformations}, in particular their effect on the key Cotton scalars~$\Psi_{\rm A}$. It is then demonstrated that a suitable null rotation can always be performed in which ${\Psi_0=0}$, identifying thus  the principle null triad and the Cotton-aligned null direction (CAND), see Sec.~\ref{CAND}. In fact, as shown in Sec.~\ref{CAND-multiplicity}, the specific  multiplicities of CANDs $\boldk$  uniquely determine the algebraic types of spacetimes. In Sec.~\ref{sec:CottonYork} we present the related symmetric traceless Cotton--York tensor, and we write it in terms of the Cotton scalars~$\Psi_{\rm A}$. Expressing it in the orthonormal basis, in Sec.~\ref{sec:CottonYorkJordan} we are able to prove a full equivalence with the previous method of classification of 2+1 geometries based on the eigenvalues and the canonical Jordan forms of the Cotton--York tensor. In Sec.~\ref{invariants} we investigate scalar curvature polynomial invariants constructed from the Cotton and Cotton--York tensors, and their relation to various algebraic types. In fact, we derive a simple practical classification algorithm based on these invariants. Sec.~\ref{sec:complexCANDs} introduces the refinement to subtypes~I$_{\rm r}$, II$_{\rm r}$, D$_{\rm r}$ for which all four (possibly multiple) CANDs are real, and subtypes~I$_{\rm c}$, II$_{\rm c}$, D$_{\rm c}$ for which  some of the CANDs are complex. This is indirectly related to complex eigenvalues of the Cotton--York tensor. In final Sec.~\ref{sec:example}, we explicitly apply this procedure on an interesting class of Robinson--Trautman spacetimes with a cosmological constant and an electromagnetic field, demonstrating that it is algebraically general (of type~I), but with only $\Psi_1$ and $\Psi_3$ scalars non-vanishing. Similarly, we analyze several other examples of metrics of various algebraic types and subtypes.

\newpage


\section{Cotton tensor}
\label{sec:Cotton}

Let $(\mathcal{M},\mathbf{g})$ be a general three-dimensional Lorentzian manifold with the metric signature $(-,+,+)$. On such a manifold, at any point we construct the \emph{basis} of the tangent space consisting of three vectors $\bolde_a$, and the cotangent space \emph{dual basis} given by three 1-forms $\boldsymbol{\omega}^a$. In local coordinates $x^{\alpha }$, these are
\begin{equation}\label{basis}
\bolde_a = e^{\alpha }_a \, \partial _{\alpha } \, \qquad \boldsymbol{\omega }^a = e_{\alpha }^a \, {\rm{d}}x^{\alpha } \, .
\end{equation}
By the Latin letters ${a, b, \ldots}$ we denote the frame (anholonomic) indices, while by the Greek letters ${\alpha , \beta , \ldots}$ we denote the coordinate (holonomic) indices. In terms of the dual basis, the line element corresponding to the \emph{metric} $g_{ab}$ is
\begin{equation}
{\rm{d}}s^2 = g_{ab} \, \boldsymbol{\omega}^a  \boldsymbol{\omega}^b \, .
\end{equation}
We also assume that the manifold is equipped with the symmetric Levi-Civita connection $\nabla$.

The role of the key geometrical object in 2+1 spaces plays the (conformally invariant) \emph{Cotton tensor}, first investigated by Cotton \cite{Cotton:1899} already in 1899, and later by Schouten \cite{Schouten}. It is the best analogue for the Weyl tensor which identically vanishes in 2+1 geometries. The Cotton tensor is defined as
\begin{equation} \label{Cotton_Definition}
C_{abc} \equiv 2\, \Big( \nabla_{[a}R_{b]c}-\frac{1}{4}\nabla_{[a}R\,g_{b]c} \Big),
\end{equation}
where $R_{ab}$ is the \emph{Ricci tensor} of the metric $g_{ab}$, see equation (20.39) in \cite{Garcia:2017}. From the definition \eqref{Cotton_Definition} it follows that the Cotton tensor is \emph{antisymmetric in the first two indices},\footnote{Unfortunately, in mathematical literature a different convention is also used for the position of the antisymmetric indices, as for example in equation (3.89) in \cite{Stephani}.}
\begin{equation} \label{anti}
C_{abc}=-C_{bac} \, ,
\end{equation}
and that it also satisfies the constraints
\begin{align}
C_{[abc]}    &=0 \, , \label{B1}\\
{C_{ab}}^{a} &=0 \, . \label{B2}
\end{align}
For a detailed exposition of the Cotton tensor see \cite{GHHM} or Chapter~20 in \cite{Garcia:2017}. These constraints restrict the Cotton tensor in 2+1 geometries to have \emph{only 5 independent components}. Indeed, due to \eqref{anti}, the Cotton tensor has ${3 \times 3 = 9}$ independent components which are constrained by 1~condition \eqref{B1} and 3~independent conditions \eqref{B2}.


\section{Null triad and the Cotton scalars~$\Psi_{\rm A}$}
\label{sec:PsiAdef}

The next step is to project the Cotton tensor onto a suitable basis on the tangent space. We choose the \emph{null triad} ${\{ \bolde_a \}\equiv\{ \boldk, \, \boldl, \, \boldm \}}$, such that $\boldk \cdot \boldk=0=\boldl \cdot \boldl$, $\boldk \cdot \boldm=0=\boldl \cdot \boldm$, and
\begin{equation} \label{null}
\boldk \cdot \boldl=-1 \, , \qquad \boldm \cdot \boldm=1 \, ,
\end{equation}
or written explicitly in the components
\begin{equation} \label{null-comp}
k_a \, l^a=-1 \, , \qquad m_a \, m^a=1 \, .
\end{equation}
It means that both  $\boldk$ and $\boldl$ are \emph{null vectors} (future-oriented and mutually normalized to $-1$), while $\boldm$ is the \emph{spatial unit vector} orthogonal to $\boldk$ and $\boldl$.

A dual basis $\{ \boldsymbol{\omega }^b\}$ is given by the relation  ${e_a^{\alpha } \, \omega ^b_{\alpha}=\delta ^b_a}$. In view of the scalar products \eqref{null}, such a dual basis can be written as $\{ \boldsymbol{\omega }^b\}\equiv\{ -\boldl, -\boldk, \boldm \}$. By this notation we mean that the dual to the vector ${\bolde_1=\boldk=k^{\alpha} \, \partial _{\alpha } }$ is the 1-form $\boldsymbol{\omega }^1=-l_{\alpha } \, {\rm{d}}x^{\alpha}$, and similarly for the remaining two basis vectors.

Now we define the \emph{Newman--Penrose-type} curvature \emph{Cotton scalars}~$\Psi_{\rm A}$ as
\begin{align}
\Psi _0 &\equiv C_{abc} \, k^a \, m^b \, k^c \, , \nonumber\\ 
\Psi _1 &\equiv C_{abc} \, k^a \, l^b \, k^c \, , \nonumber\\
\Psi _2 &\equiv C_{abc} \, k^a \, m^b \, l^c \, , \label{Psi}\\
\Psi _3 &\equiv C_{abc} \, l^a \, k^b \, l^c \, , \nonumber\\
\Psi _4 &\equiv C_{abc} \, l^a \, m^b \, l^c \, . \nonumber   
\end{align}
These are fully analogous to standard definition of the Newman--Penrose scalars constructed from the \emph{Weyl} curvature tensor in ${D=4}$ (see \cite{Stephani}) and in any ${D>4}$ (see \cite{KrtousPodolsky:2006}, equivalent to \cite{ColeyMilsovPravdaPravdova:2004, MilsonColeyPravdaPravdova:2005, Coley:2008, OrtaggioPravdaPravdova:2013}). Notice that these scalars are \emph{real}, and completely represent the \emph{5 independent components} of the Cotton tensor.


\section{Algebraic classification based on the $\Psi_{\rm A}$~scalars}
\label{sec:PsiAclassification}

We propose that the algebraic classification of 2+1 geometries can easily be made by using these curvature scalars~$\Psi_{\rm A}$, which are the components of the Cotton tensor with respect to the null triad, defined in \eqref{Psi}. The specific algebraic types are given by simple conditions, namely that \emph{in a suitable triad}
${\{ \boldk, \, \boldl, \, \boldm \}}$ the \emph{specific Cotton scalars vanish}, as summarized in Table~\ref{Tab-classification}.

\begin{table}[!h]
\begin{center}
\begin{tabular}{ c c l l}
\hline
\\[-8pt]
algebraic type && \qquad\quad the conditions\\[2pt]
\hline
\\[-8pt]
I   && ${\Psi _0=0}$\,,                              & $\Psi _1\ne0$ \\[2pt]
II  && ${\Psi _0=\Psi _1=0}$\,,                      & $\Psi _2\ne0$ \\[2pt]
III && ${\Psi _0=\Psi _1=\Psi _2=0}$\,,              & $\Psi _3\ne0$ \\[2pt]
N   && ${\Psi _0=\Psi _1=\Psi _2=\Psi _3=0}$\,,      & $\Psi _4\ne0$ \\[2pt]
D   && ${\Psi _0=\Psi _1 = 0 = \Psi _3=\Psi _4}$\,,  & $\Psi _2\ne0$ \\[2pt]
O   && all ${\Psi_{\rm A}=0}$  & \\[2pt]
\hline
\end{tabular}
\caption{The algebraic classification of 2+1 geometries.}
\label{Tab-classification}
\end{center}
\end{table}

In fact, this is a direct analogue of the Petrov--Penrose algebraic classification in standard general relativity based on the multiplicity of the PNDs of the Weyl tensor (see Section~4.3 in \cite{Stephani} for the review), or of PAT/SAT and the multiplicity of the WANDs in higher dimensions (see \cite{OrtaggioPravdaPravdova:2013}).

To justify the definition of algebraic types presented in Table~\ref{Tab-classification} and to demonstrate that it is equivalent to the previous definition based on the Jordan forms of the Cotton--York tensor, it is  now necessary to introduce a convenient bivector basis of 2-forms, which effectively represent the first two (antisymmetric) indices of the Cotton tensor \eqref{Cotton_Definition}.


\section{Cotton tensor in the bivector basis}
\label{sec:Cottonbivector}

The space of all \emph{2-forms} (also called \emph{bivectors}) in 2+1 geometries has dimension 3, and we now construct a basis
\begin{equation}
\{ \boldZ^I \}=\{\boldU, \boldV, \boldW \} \, ,
\end{equation}
where ${I=1,2,3}$, to express them. In particular, employing the null triad ${\{ \boldk, \, \boldl, \, \boldm \}}$ normalized as \eqref{null}, we define these base 2-forms as the wedge products
\begin{align}
\boldU& \equiv 2 \, \boldm\wedge \boldl \, , \nonumber\\
\boldV& \equiv 2 \, \boldk\wedge \boldm \, , \label{Vbivec}\\
\boldW& \equiv 2 \, \boldl\wedge \boldk \, . \nonumber  
\end{align}
More explicitly, in the null  triad frame such a bivector basis is
${\{ Z^I_{ab} \}=\{ U_{ab}\, ,V_{ab}, \, W_{ab}\}}$, where
\begin{align}
U_{ab}&= m_a \, l_b-l_a \, m_b \, , \nonumber\\
V_{ab}&= k_a \, m_b-m_a \, k_b \, , \label{Vbivec2}\\
W_{ab}&= l_a \, k_b-k_a \, l_b \, . \nonumber  
\end{align}
It is the analogous definition as in ${D=4}$, see Eq.~(3.40) in \cite{Stephani}. A direct calculation using  \eqref{null-comp} reveals that these bivectors satisfy the normalization relations
\begin{equation} \label{Bivector_Contraction}
U_{ab} \, V^{ab}=2 \, , \qquad W_{ab} \, W^{ab}=-2 \, ,
\end{equation}
while all other contractions are zero.

The rank-3 Cotton tensor, antisymmetric in the first two indices, can be expressed in the basis given by (all combinations of) the tensor product of a basis bivector $ \boldZ^I$ and a 1-form $\boldsymbol{\omega}^J$, that is
\begin{equation}\label{Comp-of-Cabc}
C_{abc}=\sum_{I ,J =1 }^3 C_{IJ} \, Z^I_{ab} \, \omega ^J_{c} \, ,
\end{equation}
where ${C_{IJ}}$ are the corresponding components.\footnote{Here ${Z^I_{ab} \, \omega ^J_{c}}$ is a shorthand for ${Z^I_{ab} \otimes \omega ^J_{c}}$.} Written explicitly, it has nine terms,
\begin{align}
C_{abc}=&-C_{11} \, U_{ab} \, l_c-C_{12} \, U_{ab} \, k_c+C_{13} \, U_{ab} \, m_c \nonumber \\
&-C_{21} \, V_{ab} \, l_c-C_{22} \, V_{ab} \, k_c+C_{23} \, V_{ab} \, m_c  \\
&-C_{31} \, W_{ab} \, l_c-C_{32} \, W_{ab} \, k_c+C_{33} \, W_{ab} \, m_c \, . \nonumber
\end{align}
Since the bivectors ${U_{ab}, V_{ab}, W_{ab}}$ are antisymmetric, the condition \eqref{anti} is trivially satisfied. Now we employ the vanishing trace condition \eqref{B2}. Using the relations
\begin{align}
U_{ab} \, l^a&=0 \, , & U_{ab} \, k^a&=m_b \, , & U_{ab} \, m^a&=l_b \, , \nonumber\\
V_{ab} \, l^a&=-m_b \, , & V_{ab} \, k^a&=0 \, , & V_{ab} \, m^a&=-k_b \, , \\
W_{ab} \, l^a&=l_b \, , & W_{ab} \, k^a&=-k_b \, , & W_{ab} \, m^a&=0 \, , \nonumber
\end{align}
we obtain the constraint
\begin{equation}
(C_{13}-C_{31}) \, l_b+(C_{32}-C_{23}) \, k_b+(C_{21}-C_{12}) \, m_b=0 \, .
\end{equation}
This 1-form must be identically zero, and so we obtain three conditions for the components, namely
\begin{equation} \label{Coeff_Conditions_I}
C_{13}=C_{31} \, , \qquad C_{23}=C_{32} \, , \qquad C_{12}=C_{21} \, .
\end{equation}
The Cotton tensor thus can be written in the form
\begin{align}
C_{abc}=&-C_{11} \, U_{ab} \, l_c-C_{12}(U_{ab} \, k_c+V_{ab} \, l_c)+C_{13}(U_{ab} \, m_c-W_{ab} \, l_c)  \nonumber \\
&-C_{22} \, V_{ab} \, k_c+C_{23}(V_{ab} \, m_c-W_{ab} \, k_c)+C_{33} \, W_{ab} \, m_c \, . \label{1st_General_Cotton}
\end{align}

Finally, we have to apply the remaining condition \eqref{B1}. It is helpful first to calculate that
\begin{align}
3! \, (U_{[ab} \, k_{c]}+V_{[ab} \, l_{c]})& = 4\,(U_{ab} \, k_c+V_{ab} \, l_c+W_{ab} \, m_c) \, ,\nonumber\\
3! \, W_{[ab} \, m_{c]}& = 2\,(U_{ab} \, k_c+V_{ab} \, l_c+W_{ab} \, m_c) \, ,
\end{align}
where the factorial was included just to compensate the factor in the definition of the antisymmetrization. All other terms of the tensor-product basis are trivially zero under the complete antisymmetrization, namely
\begin{align}
U_{[ab} \, l_{c]}&=0 \, , & U_{[ab} \, m_{c]}&=0 \, , \nonumber\\
V_{[ab} \, k_{c]}&=0 \, , & V_{[ab} \, m_{c]}&=0 \, ,\\
W_{[ab} \, k_{c]}&=0 \, , & W_{[ab} \, l_{c]}&=0 \, . \nonumber
\end{align}
From the condition \eqref{B1} for \eqref{1st_General_Cotton} we now obtain
\begin{equation}
(-2C_{12}+C_{33})(U_{ab} \, k_c+V_{ab} \, l_c+W_{ab} \, m_c)=0\, ,
\end{equation}
which implies the last constraint
\begin{equation} \label{Coeff_Conditions_II}
C_{33}=2 \, C_{12}\, .
\end{equation}
The \emph{generic Cotton tensor} in the bivector-null basis thus takes the form
\begin{align}
C_{abc}=&-C_{11} \, U_{ab} \, l_c-C_{12}(U_{ab} \, k_c+V_{ab} \, l_c-2W_{ab} \, m_c)  \nonumber \\
&+C_{13}(U_{ab} \, m_c-W_{ab} \, l_c)-C_{22} \, V_{ab} \, k_c+C_{23}(V_{ab} \, m_c-W_{ab} \, k_c) \, . \label{2nd_General_Cotton}
\end{align}
It has 5 independent components, namely $C_{11}, C_{12}, C_{13}, C_{22} , C_{23}$. They can be \emph{uniquely} expressed in terms of the Newman--Penrose-type curvature \emph{Cotton scalars}~$\Psi_{\rm A}$ defined in \eqref{Psi}. Indeed, using the normalization relations \eqref{null-comp} and \eqref{Bivector_Contraction}, the coefficients in \eqref{2nd_General_Cotton} can be expressed as
\begin{alignat}{3}
C_{11}&=\tfrac{1}{2}C_{abc} \, V^{ab} \, k^c \, , \nonumber\\
C_{12}&=\tfrac{1}{2}C_{abc} \, V^{ab} \, l^c &&=\tfrac{1}{2}C_{abc} \, U^{ab} \, k^c
   &&\hspace{-4mm}=-\tfrac{1}{4}C_{abc} \, W^{ab} \, m^c \, , \nonumber\\
C_{13}&=\tfrac{1}{2}C_{abc} \, V^{ab} \, m^c & &= -\tfrac{1}{2}C_{abc} \, W^{ab} \, k^c \, , \label{Coeff_Cab}\\
C_{22}&=\tfrac{1}{2}C_{abc} \, U^{ab} \, l^c \, , \nonumber\\
C_{23}&=\tfrac{1}{2}C_{abc} \, U^{ab} \, m^c & &= -\tfrac{1}{2}C_{abc} \, W^{ab} \, l^c \, . \nonumber 
\end{alignat}
After explicitly putting the bivectors \eqref{Vbivec2} into the first terms on the right-hand side of \eqref{Coeff_Cab}, using the definition \eqref{Psi} of the scalars~$\Psi_{\rm A}$ and the antisymmetry of the Cotton tensor \eqref{anti}, we arrive at a very simple expressions for the five independent components of the Cotton tensor, namely
\begin{equation} \label{coeff}
C_{11}=\Psi _0 \, ,\quad C_{12}=\Psi _2 \, ,\quad C_{13}=\Psi _1 \, ,\quad C_{22}=-\Psi _4 \, ,\quad C_{23}=-\Psi _3 \, .
\end{equation}
Moreover, the four other basis components of the Cotton tensor in the expansion \eqref{Comp-of-Cabc} are \emph{not} independent because from the remaining four expressions on the right-hand side of \eqref{Coeff_Cab} we get
\begin{alignat}{2}
C_{12}&=C_{abc} \, m^a \, l^b \, k^c & =\tfrac{1}{2}C_{abc} \, k^a \, l^b \, m^c \, , \nonumber\\
C_{13}&=C_{abc} \, k^a \, m^b \, m^c \, ,\\
C_{23}&=C_{abc} \, m^a \, l^b \, m^c \, .\nonumber
\end{alignat}
Using \eqref{coeff} we can thus write that
\begin{alignat}{2}
\Psi_1&=C_{abc} \, k^a \, m^b \, m^c \, ,\nonumber\\
\Psi_2&=C_{abc} \, m^a \, l^b \, k^c & =\tfrac{1}{2}C_{abc} \, k^a \, l^b \, m^c \, , \label{Psi2-alter}\\
\Psi_3&=C_{abc} \, l^a \, m^b \, m^c \, ,\nonumber
\end{alignat}
which are the four \emph{alternative} expressions for the three scalars $\Psi_1, \Psi_2, \Psi_3$, equivalent to their definitions given in \eqref{Psi}.

We can thus conclude that the \emph{most general Cotton tensor} in the bivector-null basis \eqref{2nd_General_Cotton} takes the form
\begin{align}
C_{abc}=&-\Psi _0 \, U_{ab} \, l_c+\Psi _1 \,(U_{ab} \, m_c-W_{ab} \, l_c) \nonumber \\
&-\Psi _2 \,(U_{ab} \, k_c+V_{ab} \, l_c-2W_{ab} \, m_c) \label{Cotton_General}\\
&-\Psi _3 \,(V_{ab} \, m_c-W_{ab} \, k_c)+\Psi _4 \, V_{ab} \, k_c \, . \nonumber
\end{align}
This is an important expression of the Cotton tensor in terms of the five key scalars~$\Psi_{\rm A}$, which we will employ in proving many further properties and relations. In fact, it is obviously an analogue of the standard expression valid in ${D=4}$ general relativity, see Eq.~(3.58) in \cite{Stephani}.


\section{Bel--Debever criteria}
\label{Bel-Debever-criteria}

It is now possible to explicitly connect the algebraic classification of 2+1 gravity fields, summarized in Table~\ref{Tab-classification}, to another property, namely to the \emph{Bel--Debever criteria} which involve the Cotton tensor and the related \emph{(aligned) null vectors}~$\boldk$.

These criteria were presented in 1959 by Bel and Debever \cite{Bel:1959, Debever:1959a, Debever:1959b} in the context of Einstein's general relativity in ${D=4}$, employing the Riemann or Weyl tensors and the corresponding Debever--Penrose null vectors $\boldk$. Relatively recently, they were also generalized to geometries of any higher dimension ${D>4}$ by Ortaggio \cite{Ortaggio:2009} (using the principal directions of the Weyl tensor) and equivalently by Senovilla  \cite{Senovilla:2010} (using the Bel--Robinson tensor).

We claim that in ${D=3}$ spacetimes the Bel--Debever criteria involve the Cotton scalars, and they have the following form:
\begin{align}
k_{[d}\,C_{a]bc} \, k^b \, k^c &=0 \quad \Leftrightarrow \quad \Psi _0=0 \, , \label{BD-I}\\
C_{abc} \, k^b \, k^c &=0 \quad \Leftrightarrow \quad \Psi _0=\Psi _1=0 \, ,  \label{BD-II}\\
k_{[d}\,C_{a]bc} \, k^b &=0 \quad \Leftrightarrow \quad \Psi _0=\Psi _1=\Psi _2=0 \, , \label{BD-III}\\
C_{abc} \, k^b &=0 \quad \Leftrightarrow \quad \Psi _0=\Psi _1=\Psi _2=\Psi _3=0 \, . \label{BD-N}
\end{align}

The proof is not difficult. Using \eqref{Cotton_General} we get
\begin{equation} \label{last}
C_{abc} \, k^b=\Psi _0 \, m_a \, l_c-\Psi _1 \, (m_a \, m_c+k_a \, l_c)+\Psi _2 \, (m_a \, k_c +2k_a \, m_c)+\Psi _3 \, k_a \, k_c \, ,
\end{equation}
and then
\begin{equation} \label{equiv}
C_{abc} \, k^b  \, k^c=-\Psi _0 \, m_a+\Psi _1 \, k_a \, .
\end{equation}
After multiplying this expression by $k_d$, the antisymmetrization yields
\begin{equation}
k_{[d}\,C_{a]bc} \, k^b \, k^c = \tfrac{1}{2}\,\Psi _0 \, V_{a d} \, ,
\end{equation}
from which we obtain the equivalence \eqref{BD-I}. The equivalence \eqref{BD-II} follows immediately from \eqref{equiv}.
As for \eqref{BD-III}, we employ \eqref{last} which implies
\begin{equation}
k_{[d}\,C_{a]bc} \, k^b = \tfrac{1}{2}\, V_{d a} \,( \Psi _0 \, l_c - \Psi _1 \, m_c + \Psi _2 \,k_c )\, ,
\end{equation}
from which the equivalence \eqref{BD-III} is clear. The last relation \eqref{BD-N} is obvious from \eqref{last}.

For type D spacetimes, not only ${\Psi _0=\Psi _1 = 0}$ but also ${\Psi _4=\Psi _3=0}$. Because
\begin{equation}
C_{abc} \, l^b \, l^c=\Psi _3 \, l_a -\Psi _4 \, m_a \, ,
\end{equation}
it follows that
\begin{equation}
C_{abc} \, l^b \, l^c =0 \quad \Leftrightarrow \quad \Psi _3=\Psi _4=0 \, .
\end{equation}
Therefore, for type D geometries \emph{both} ${C_{abc} \, k^b  \, k^c = 0}$ and ${C_{abc} \, l^b  \, l^c = 0}$. This concludes the proof of the Bel--Debever criteria in 2+1 gravity.

The results for all algebraic types are summarized in Table~\ref{Tab:DB-criteria}. The second column contains the Bel--Debever criteria in 2+1 geometries, while the last column contains the classic Bel--Debever criteria in 3+1 geometries, see Eqs.~(4.21)--(4.24) and (4.27) in \cite{Stephani}. It is obvious that there is a perfect analogy when the Weyl tensor $C_{abcd}$ is replaced by the Cotton tensor $C_{abc}$ in lower dimension ${D=3}$.

\begin{table}[!h]
\begin{center}
\begin{tabular}{crr}
\hline
\\[-8pt]
   algebraic type & \quad 2+1 geometries & 3+1 geometries   \\[2pt]
\hline
\\[-8pt]
   I
   & ${k_{[d}\,C_{a]bc} \, k^b \, k^c =0}$
   & ${\qquad k_{[f}\,C_{a]bc [d} \,k_{e ]}\,k^b\, k^c =0}$ \\[4pt]
   II
   & ${C_{abc} \, k^b \, k^c =0}$
   & ${C_{abc [d}\,k_{e ]}k^b\, k^c =0}$ \\[4pt]
   III
   & ${k_{[d}\,C_{a]bc} \, k^b =0}$
   & ${k_{[f}\,C_{a]bcd}\, k^b =0}$ \\[4pt]
   N
   & ${C_{abc} \, k^b =0}$
   & ${C_{abcd}\, k^b =0}$  \\[4pt]
   D
   & ${C_{abc} \, k^b \, k^c =0}$
   & ${C_{abc [d}\,k_{e ]}\,k^b\, k^c =0}$ \\[0pt]
   & and\ ${C_{abc} \, l^b \, l^c =0}$
   & and\ ${C_{abc [d}\,l_{e ]}\,l^b\, l^c =0}$  \\[4pt]
   O
   & ${C_{abc} =0 }$
   & ${C_{abcd}=0 }$ \\[2pt]
\hline
\end{tabular}
\caption{\label{Tab:DB-criteria} For all algebraic types, the Bel--Debever criteria in 2+1 geometries involving the Cotton tensor $C_{abc}$ (left) and in 3+1 geometries involving the Weyl tensor $C_{abcd}$ (right) are fully analogous.}
\end{center}
\end{table}

In fact, the privileged (aligned) null vectors $\boldk$ which enter the Bel--Debever criteria in Table~\ref{Tab:DB-criteria} are the (possibly multiple) \emph{principal null directions} of the Cotton and the Weyl tensor, respectively. Now we will demonstrate that these can be systematically investigated and easily found also by using the Newman--Penrose-type Cotton scalars~$\Psi_{\rm A}$ defined in \eqref{Psi}.


\section{Lorentz transformations of the Cotton scalars~$\Psi_{\rm A}$}
\label{Lorentz-transformations}

The key curvature scalars~$\Psi_{\rm A}$, which conveniently represent five independent components of the Cotton tensor \eqref{Psi}, are not unique in the sense that they \emph{depend on the choice of the null triad} ${\{ \boldk, \, \boldl, \, \boldm \}}$. However, as in the case ${D \ge 4}$ this freedom is simple, given just by the local Lorentz transformations between various triads at a given point of the spacetime manifold. These are the only admitted changes of the null basis of the tangent space which keep the normalization conditions~\eqref{null}.

In particular, there are 3 subgroups of such Lorentz transformations, namely:
\begin{align}
\boldk' & = B\,\boldk \, ,\qquad
  \boldl' = B^{-1}\, \boldl  \, ,\qquad\,
  \boldm' = \boldm \, , \label{boost} \\[5pt]
\boldk' & = \boldk \, , \qquad
  \boldl' = \boldl + \sqrt2\, L\,\boldm + L^2\, \boldk  \, ,\qquad\,
  \boldm' = \boldm + \sqrt2\, L\,\boldk \, , \label{kfixed} \\[5pt]
\boldk' & = \boldk + \sqrt2\, K\,\boldm + K^2\, \boldl  \, ,\qquad
  \boldl' = \boldl \, ,\qquad
  \boldm' = \boldm + \sqrt2\, K\,\boldl  \, .\label{lfixed}
\end{align}
The \emph{boost} \eqref{boost} in the ${\boldk-\boldl}$ subspace is parameterized by $B$, the \emph{null rotation \eqref{kfixed} with ${\boldk}$ fixed} (changing ${\boldl}$ and ${\boldm}$) is parameterized by~$L$, while the complementary \emph{null rotation \eqref{lfixed} with ${\boldl}$ fixed} (changing~${\boldk}$ and ${\boldm}$) is parameterized by~$K$. All these three parameters ${B, K, L}$ are \emph{real}.

It is now straightforward to determine the transformation properties of the Cotton scalars \eqref{Psi}. In particular, under the \emph{boost} \eqref{boost} they  transform as a \emph{rescaling}
\begin{equation}\label{eq:boost-of-Psi}
\Psi'_{\rm A} = B^{2-{\rm A}}\,\Psi_{\rm A}\, .
\end{equation}
It means that they are naturally \emph{ordered} in the definition \eqref{Psi} according to their specific \emph{boost weight}, which is the corresponding \emph{power}~${(2-{\rm A})}$ of the boost parameter~$B$. This fact is fundamental for the algebraic classification of the Cotton tensor, as an application of a general scheme developed in \cite{MilsonColeyPravdaPravdova:2005, OrtaggioPravdaPravdova:2013} for any tensor.

Under the null rotation \eqref{kfixed} the Cotton scalars transform as
\begin{align}\label{eq:null-rotatin-fixed-k}
\Psi_0' &= \Psi_0 \, , \nonumber\\
\Psi_1' &= \Psi_1 + \sqrt{2}\,L\,\Psi_0 \, , \nonumber\\
\Psi_2' &= \Psi_2 + \sqrt{2}\,L\,\Psi_1 + L^2\,\Psi_0 \, ,\\
\Psi_3' &= \Psi_3 -3\sqrt{2}\,L\,\Psi_2 -3L^2\,\Psi_1 - \sqrt{2}\,L^3\,\Psi_0 \, , \nonumber\\
\Psi_4' &= \Psi_4 +2\sqrt{2}\,L\,\Psi_3 -6L^2\,\Psi_2 -2\sqrt{2}\,L^3\,\Psi_1-L^4\, \Psi_0 \, . \nonumber
\end{align}

It follows that the classification of 2+1 geometries summarized in Table~\ref{Tab-classification} is \emph{invariant with respect to both types of the Lorentz transformations \eqref{boost} and \eqref{kfixed}}. Indeed, if the corresponding condition for a certain algebraic type is satisfied for~$\Psi_{\rm A}$, it remains satisfied for~$\Psi_{\rm A}'$.

Finally --- and more importantly --- it remains to investigate the effect of the null rotation \eqref{lfixed} \emph{with fixed} ${\boldl'=\boldl}$ which changes the vectors $\boldk$ and $\boldm$ of the null triad to $\boldk'$ and $\boldm'$. In such a case the Cotton scalars \eqref{Psi} transform as
\begin{align}\label{eq:null-rotatin-fixed-l}
\Psi_0' &= \Psi_0+2\sqrt{2}\,K\,\Psi_1 +6K^2\,\Psi_2 -2\sqrt{2}\,K^3\,\Psi_3 - K^4\,\Psi_4 \, , \nonumber\\
\Psi_1' &= \Psi_1+3\sqrt{2}\,K\,\Psi_2 -3K^2\,\Psi_3 - \sqrt{2}\,K^3\,\Psi_4 \, , \nonumber\\
\Psi_2' &= \Psi_2 -\sqrt{2}\,K\,\Psi_3 - K^2\,\Psi_4\, ,\\
\Psi_3' &= \Psi_3 +\sqrt{2}\,K\,\Psi_4 \, , \nonumber\\
\Psi_4' &= \Psi_4 \, . \nonumber
\end{align}
Notice that these expressions are complementary to \eqref{eq:null-rotatin-fixed-k} under the swap of the null vectors ${\boldk \leftrightarrow \boldl}$ and ${K \leftrightarrow L}$, which implies ${\Psi_0 \leftrightarrow \Psi_4 }$, ${\Psi_1 \leftrightarrow \Psi_3}$ and
${\Psi_2 \leftrightarrow -\Psi_2 }$, see \eqref{Psi} and  \eqref{Psi2-alter}.


\section{Principle null triad and the Cotton-aligned null direction (CAND)}
\label{CAND}

Now we come to a crucial observation, namely that the null rotation \eqref{eq:null-rotatin-fixed-l} \emph{always} allows us to \emph{achieve}~${\Psi'_0=0}$ by a \emph{suitable choice} of the (complex) parameter $K$. Consequently, in the new null triad  ${\{ \boldk', \, \boldl', \, \boldm' \}}$ the condition for algebraic type~I given in Table~\ref{Tab-classification} is satisfied. Such a special frame is called the \emph{principle null triad}, and its special null vector $\boldk'$ is said to be \emph{aligned with the Cotton curvature tensor} $C_{abc}$. The existence of the  principal null triad demonstrates that \emph{all 2+1 geometries are (at least) of algebraic type~I}. Recall that the same is true for all ${3+1}$ geometries, considering the Weyl tensor instead of the Cotton tensor, but in higher-dimensional spacetimes such a principal null frame need not exist at all (see \cite{Stephani} and \cite{OrtaggioPravdaPravdova:2013}, respectively).

For practical reasons, however, we will consider the equivalent opposite procedure, in which one starts with the Cotton scalars $\Psi_{\rm A}'$ calculated with respect to an \emph{arbitrarily chosen} null triad ${\{ \boldk', \, \boldl', \, \boldm' \}}$. It is then \emph{possible to achieve}~${\Psi_0=0}$ by performing the \emph{inverse} of the null rotation \eqref{lfixed}, that is
\begin{eqnarray}
\boldk = \boldk' - \sqrt2\, K\,\boldm' + K^2\, \boldl'  \, ,\qquad
  \boldl = \boldl' \, ,\qquad
  \boldm = \boldm' - \sqrt2\, K\,\boldl'  \, .
 \label{tetradtrans-repeated}
\end{eqnarray}
The resulting special triad ${\{ \boldk, \, \boldl, \, \boldm \}}$ becomes the  \emph{principle null triad}, and its null vector $\boldk$ is the \emph{Cotton-aligned null direction}, which we can abbreviate as CAND. It is the 2+1 analogue of the usual concept of PND (principal null direction) of the Weyl tensor in ${D=4}$ general relativity, and of WAND (Weyl-aligned null direction) in  ${D\ge4}$  gravity, as introduced in \cite{ColeyMilsovPravdaPravdova:2004, MilsonColeyPravdaPravdova:2005}.

In fact, such an algebraically privileged triad with the CAND can be \emph{explicitly found}. Under the null rotation \eqref{tetradtrans-repeated} the Cotton scalar $\Psi_0$ (having the highest boost weight $+2$) transforms as
\begin{align}\label{Psi0:null-rotatin-fixed-l}
\Psi_0 = \Psi_0'-2\sqrt{2}\,K\,\Psi_1' +6K^2\,\Psi_2' +2\sqrt{2}\,K^3\,\Psi_3' - K^4\,\Psi_4' \, .
\end{align}
Actually, it is obtained from \eqref{eq:null-rotatin-fixed-l} by the simple swap ${\Psi_{\rm A} \leftrightarrow \Psi_{\rm A}'}$ and ${K \leftrightarrow -K}$. The condition ${\Psi_0=0}$ thus takes the form
\begin{equation} \label{Class_eq}
\Psi_4'\,K^4 - 2\sqrt{2}\,\Psi_3'\,K^3 - 6\,\Psi_2'\,K^2 + 2\sqrt{2}\,\Psi_1'\,K  - \Psi_0' =0 \, .
\end{equation}
It is an \emph{algebraic} equation of the \emph{fourth order} in the parameter $K$ which, in general, admits \emph{four complex solutions}  (not necessarily distinct). It thus follows that, \emph{at any event} of the 2+1 spacetime there exist, in general, \emph{four CANDs} determined by the local algebraic structure of the (non-vanishing) Cotton tensor.

Each of these Cotton-aligned null directions~$\boldk$ is obtained using the relation \eqref{tetradtrans-repeated}, in which the parameter $K$ is the corresponding root of the equation \eqref{Class_eq}. Moreover, any  multiplicity of these roots $K$ implies the same \emph{multiplicity of the CANDs}. We will now demonstrate that such multiplicities are uniquely related to the algebraic types.

\begin{table}[t]
\begin{center}
\begin{tabular}{cccll}
\hline
\\[-8pt]
  algebraic type & CANDs & {multiplicity\quad} & & \hspace{-36mm} canonical Cotton scalars \\[2pt]
\hline
\\[-8pt]
   I
   & \hbox{
   \rotatebox[origin=c]{-30}{$\leftarrow$}\hspace{-3mm}
   \raisebox{1.5mm}{\rotatebox[origin=c]{-60}{$\leftarrow$}}\hspace{-1mm}
   \raisebox{1.5mm}{\rotatebox[origin=c]{60}{$\rightarrow$}}\hspace{-3mm}
   \rotatebox[origin=c]{30}{$\rightarrow$}}
   & \hspace{-1mm}${1+1+1+1\quad}$
   & ${\Psi_0=0}$\,,
   & ${\Psi_1\ne0}$     \\[2pt]
   II
   & \hbox{
   \rotatebox[origin=c]{-30}{$\leftarrow$}\hspace{-3mm}
   \raisebox{1.5mm}{\rotatebox[origin=c]{-60}{$\leftarrow$}}\hspace{-1mm}
   \raisebox{0.6mm}{\rotatebox[origin=c]{45}{$\Rightarrow$}}}
   & ${1+1+2\quad}$
   & ${\Psi_0=\Psi_1=0}$\,,
   & ${\Psi_2\ne0}$    \\[6pt]
   D
   & \hbox{
   \rotatebox[origin=c]{-45}{$\Leftarrow$}\,\rotatebox[origin=c]{45}{$\Rightarrow$}}
   & ${2+2\quad}$
   & ${\Psi_0=\Psi_1=0=\Psi_3=\Psi_4}$\,,
   & ${\Psi_2\ne0}$    \\[4pt]
   III
   & \hbox{
   \rotatebox[origin=c]{-30}{$\leftarrow$}\hspace{-1mm}
   \raisebox{0.6mm}{\rotatebox[origin=c]{45}{$\Rrightarrow$}}}
   & ${1+3\quad}$
   & ${\Psi_0=\Psi_1=\Psi_2=0}$\,,
   & ${\Psi_3\ne0}$     \\[6pt]
   N
   & {\Large
   \hbox{
   \rotatebox[origin=c]{45}{$\Rightarrow$}\hspace{-6.1mm}
   \raisebox{-0.30mm}{\rotatebox[origin=c]{45}{$\Rightarrow$}}}
   }
   & $4\quad$
   & ${\Psi_0=\Psi_1=\Psi_2=\Psi_3=0}$\,,
   & ${\Psi_4\ne0}$     \\[6pt]
   O
   &
   &  {N/A\quad}
   & all ${\Psi_{\rm A}=0}$
   &    \\[2pt]
\hline
\end{tabular}
\caption{Possible algebraic types of 2+1 geometries. The classification is uniquely related to the multiplicity of the Cotton-aligned null directions (CANDs), that is to the multiplicity of the four (complex) roots of the key equation \eqref{Class_eq}. The canonical forms of the five real Cotton scalars~$\Psi_{\rm A}$ for each algebraic type are also included.}
\label{Tab:algebraic-types}
\end{center}
\end{table}


\section{Algebraic types and the CANDs multiplicity}
\label{CAND-multiplicity}

A 2+1 spacetime is said to be \emph{algebraically general} if its CANDs, i.e. the four roots of \eqref{Class_eq}, are \emph{all distinct}. Such a spacetime is of algebraic \emph{type~I}.

A spacetime is \emph{algebraically special} if at least two its CANDs \emph{coincide}. If \emph{just two} CANDs~$\boldk$ coincide, it is of \emph{type~II}. Analogously, higher multiplicity defines \emph{type~III} (triple CAND/root) and the most special \emph{type~N} (quadruple CAND/root) geometries.

In addition, there exists another degenerate case of \emph{type~D}. It is a subtype of type~II such that there are two distinct CANDs~$\boldk$ and~$\boldl$, both of multiplicity 2 (\emph{two pairs} of coinciding roots). For completeness, the algebraic type~O denotes a spacetime with everywhere vanishing Cotton tensor (a \emph{conformally flat} 2+1 spacetime). The complete scheme is summarized in Table~\ref{Tab:algebraic-types}.

More specifically, if the vector $\boldk$ of the principal null triad is the Cotton-aligned null direction then ${\Psi_0=0}$, and the key equation \eqref{Class_eq} in such a triad becomes
\begin{equation} \label{Class_eq-special}
(\Psi_4\,K^3 - 2\sqrt{2}\,\Psi_3\,K^2 - 6\,\Psi_2\,K + 2\sqrt{2}\,\Psi_1 )\,K  =0 \, .
\end{equation}
The root ${K=0}$ corresponds to the CAND $\boldk$. The special algebraic types arise when also the cubic expression in the bracket  has another root(s) ${K=0}$. It is now obvious that type~II arises when ${\Psi_1=0}$, and type~III arises when ${\Psi_1=\Psi_2=0}$. Type~N occurs when ${\Psi_1=\Psi_2=\Psi_3=0}$, in which case \eqref{Class_eq-special} reduces to ${\Psi_4\,K^4  = 0}$. The quadruple root ${K=0}$ corresponds to the unique and privileged quadruple CAND $\boldk$.
For type~D spacetimes with the Cotton scalars having the form ${\Psi_0=\Psi_1=0=\Psi_3=\Psi_4}$ the equation \eqref{Class_eq-special} reduces to quadratic equation ${\Psi_2\, K^2 = 0}$, so that $\boldk$ is the double CAND, as in type~II. To be more specific:

\begin{itemize}

\item {\bf Type I} geometries with the CAND $\boldk$ satisfy the equation \eqref{Class_eq-special} in which the simple root ${K=0}$ corresponds to $\boldk$. Because ${\Psi_1 \neq 0}$ the remaining part of the equation is of the third order, admitting \emph{in general} three distinct (complex) roots different from zero. In such a generic case, there exist \emph{four different} CANDs with no multiplicities, symbolically denoted as ${1+1+1+1}$.
\vspace{2mm}

\item {\bf Type II} geometries have the canonical form ${\Psi_0=0=\Psi _1}$ and ${\Psi_2 \neq 0}$, in which case \eqref{Class_eq-special} reduces to
\begin{equation} \label{Class_eq-special-II}
(\Psi_4\,K^2 - 2\sqrt{2}\,\Psi_3\,K - 6\,\Psi_2)\,K^2  =0 \, .
\end{equation}
It has the solution ${K=0}$ with the \emph{multiplicity $2$} (which means that $\boldk$ is a \emph{double} CAND), and other ${K\neq0}$ solutions are given by the roots of the quadratic equation in the bracket. In general, there exists two different (complex) roots, meaning that the multiplicities of the \emph{three different} CANDs are ${1+1+2}$.

\vspace{2mm}

\item {\bf Type III} geometries have the canonical form of the Cotton scalars ${\Psi_0=\Psi_1=\Psi_2=0}$ with ${\Psi_3\ne0}$. The key equation \eqref{Class_eq-special} thus takes the form
\begin{equation} \label{Class_eq-special-III}
(\Psi_4\,K - 2\sqrt{2}\,\Psi_3)\,K^3  =0 \, .
\end{equation}
The trivial solution ${K=0}$ has the \emph{multiplicity $3$} (which means that $\boldk$ is a \emph{triple} CAND), and there exists another (real) root  ${K = 2\sqrt{2}\,\Psi_3 / \Psi_4 \neq0}$. The multiplicities of the \emph{two different} CANDs are thus ${1+3}$.

\vspace{2mm}

\item {\bf Type N} geometries are defined by the canonical condition ${\Psi_4\ne0}$ with all remaining Cotton scalars zero, so that the equation \eqref{Class_eq-special} simplifies to
\begin{equation} \label{Class_eq-special-N}
K^4  =0 \, .
\end{equation}
Therefore, $\boldk$ is a \emph{quadruple} CAND, corresponding to the \emph{multiplicity $4$} of the solution ${K=0}$.

\vspace{2mm}

\item {\bf Type D} geometries have the canonical form ${\Psi_0=\Psi_1=0=\Psi_3=\Psi_4}$ and ${\Psi_2\ne0}$. The key equation \eqref{Class_eq-special} thus reduces to
\begin{equation} \label{Class_eq-special-D}
K^2  =0 \, ,
\end{equation}
from which it follows that $\boldk$ is a \emph{double} CAND, corresponding to the \emph{multiplicity $2$} of the solution ${K=0}$.
In Sec.~\ref{CAND} we have defined the CAND $\boldk'$ as the null vector of the principle null triad ${\{ \boldk', \, \boldl', \, \boldm' \}}$ in which ${\Psi'_0=0}$. Analogously,\footnote{Recall that the swap of the null vectors ${\boldk \leftrightarrow \boldl}$ results in ${\Psi_0 \leftrightarrow \Psi_4 }$, ${\Psi_1 \leftrightarrow \Psi_3}$, ${\Psi_2 \leftrightarrow -\Psi_2 }$.} the null vector $\boldl'$ of the principle null triad is CAND if ${\Psi'_4=0}$.
In view of the transformation property  \eqref{eq:null-rotatin-fixed-k} of the Cotton scalars under the null rotation \eqref{kfixed} with ${\boldk}$ fixed, that is
\begin{align}
\boldk' & = \boldk \, , \qquad
  \boldl' = \boldl + \sqrt2\, L\,\boldm + L^2\, \boldk  \, ,\qquad\,
  \boldm' = \boldm + \sqrt2\, L\,\boldk \, , \label{kfixed-repeated}
\end{align}
for the canonical form of type~D geometries we obtain
\begin{align}\label{eq:null-rotatin-fixed-k-repeated}
\Psi_0' = 0 \, , \qquad
\Psi_1' = 0 \, , \qquad
\Psi_2' = \Psi_2  \, ,\qquad
\Psi_3' =  -3\sqrt{2}\,L\,\Psi_2  \, , \qquad
\Psi_4' =  -6L^2\,\Psi_2  \, .
\end{align}
Therefore, the condition ${\Psi'_4=0}$ for $\boldl'$ being the Cotton-aligned null direction is simply
\begin{equation} \label{Class_eq-special-D-l}
L^2  =0 \, .
\end{equation}
It means that the null vector ${\boldl=\boldl'}$ is, in fact, a \emph{double} CAND, corresponding to the \emph{multiplicity $2$} of the solution ${L=0}$. To summarize, geometries of algebraic type~D  admit \emph{two distinct} CANDs~$\boldk$ and~$\boldl$, both of multiplicity $2$.

\end{itemize}

This completes the proof of the relations contained in Table~\ref{Tab:algebraic-types}.

\vspace{2mm}

A special situation ${\Psi _4 =0 =\Psi _0}$ has to be treated separately. In such a case the quartic equation \eqref{Class_eq-special} reduces to the cubic
\begin{equation} \label{Class_eq-special-Invariants}
(\sqrt{2}\,\Psi_3\,K^2 + 3\,\Psi_2\,K - \sqrt{2}\,\Psi_1 )\,K  =0 \, ,
\end{equation}
with the CAND $\boldk$ (corresponding to ${K=0}$) and the distinct CAND $\boldl$ (corresponding to ${L=0}$). The respective multiplicities of the remaining roots of \eqref{Class_eq-special-Invariants} are given by the nature of the quadratic polynomial in the bracket, depending on the (non-)vanishing of the scalars $\Psi_1, \Psi_2, \Psi _3$, and also on the discriminant ${D = 9\, \Psi _2^2 +8\, \Psi _1\Psi _3}$. Full discussion of all possible algebraic types in this case is presented in Table~\ref{Tab:Algorithm-special-case}.

\newpage

\begin{table}[h!]
\begin{center}

\begin{tabular}{ c| c| c| c| l }
\hline \\[-12pt]

& \multicolumn{1}{c|}{} & \multicolumn{2}{c|}{}\\[-6pt]
\multirow{5}{*}{${\Psi _1 =0}$}  & \multirow{2}{*}{${\Psi _2 =0}$} & \multicolumn{2}{c|}{${\Psi _3 =0}$}
& type O\\[2pt] \cline{3-5}
& \multicolumn{1}{c|}{} & \multicolumn{2}{c|}{}\\[-8pt]
                              &                               & \multicolumn{2}{c|}{$\Psi _3 \neq 0$} & type III\\[2pt] \cline{2-5}
& \multicolumn{1}{c|}{} & \multicolumn{2}{c|}{}\\[-8pt]
                              & \multirow{2}{*}{$\Psi _2 \neq 0$} & \multicolumn{2}{c|}{$\Psi _3 =0$} & type D\\[2pt] \cline{3-5}
& \multicolumn{1}{c|}{} & \multicolumn{2}{c|}{}\\[-8pt]
                              &                               & \multicolumn{2}{c|}{$\Psi _3 \neq 0$} & type II \\[2pt] \hline
& \multicolumn{1}{c|}{} & \multicolumn{2}{c|}{}\\[-8pt]
\multirow{7}{*}{$\Psi _1 \neq 0$} & \multirow{2}{*}{$\Psi _2 =0$} & \multicolumn{2}{c|}{$\Psi _3 =0$} & type III\\[2pt] \cline{3-5}
& \multicolumn{1}{c|}{} & \multicolumn{2}{c|}{}\\[-8pt]
                                  &                               & \multicolumn{2}{c|}{$\Psi _3 \neq 0$} & type I\\[2pt] \cline{2-5}
& \multicolumn{1}{c|}{} & \multicolumn{2}{c|}{}\\[-8pt]
                                  & \multirow{4}{*}{$\Psi _2 \neq 0$} & \multicolumn{2}{c|}{$\Psi _3 =0$} & type II\\[2pt] \cline{3-5}
& \multicolumn{1}{c|}{} & \multicolumn{1}{c|}{} & \multicolumn{1}{c|}{}\\[-8pt]
                                  &                               & \multirow{2}{*}{$\Psi _3 \neq 0$} & $9\Psi                         																			 _2^2 =-8\Psi _1\Psi _3$ & type II\\[2pt] \cline{4-5}
& \multicolumn{1}{c|}{} & \multicolumn{1}{c|}{} & \multicolumn{1}{c|}{}\\[-8pt]
                                  &                               &                                   & $9\Psi                         																			 _2^2 \neq -8\Psi _1\Psi _3$ & type I \\[2pt]
\hline

\end{tabular}
\caption{\label{Tab:Algorithm-special-case} Algebraic classification of 2+1 geometries for the special case ${\Psi_4 =0 =\Psi _0}$.}
\end{center}
\end{table}

\vspace{2mm}

We should emphasize that the algebraic classification in 2+1 dimensions (presented here) is actually more subtle than in 3+1 geometries. The complication arises from the fact that the key \emph{real} equation \eqref{Class_eq} can in general have some \emph{complex roots} $K$. This implies that some of the null vectors $\boldk$ representing the CANDs may \emph{formally be complex}. This somewhat unwelcome consequence is closely related to the property that the important Cotton--York matrix ${Y_a}^b$, see \eqref{class} for its explicit form, \emph{is not symmetric} and there is thus \emph{no guarantee that its eigenvalues are real}. Nevertheless, it is a common practice in the field of algebraic classification of 2+1 spacetimes to \emph{formally admit} the complex classification. A geometrically more justified (sub)classification based on the \emph{real roots} can be introduced. By restricting the eigenvalues to only real numbers, a new algebraic subtype of spacetimes denoted as Class~$\text{I}^{\prime }$ can be added, see for example Sec.~20.5.2 in the Garc\'ia-D\'iaz monograph \cite{Garcia:2017}. We will return to this issue later in Sec.~\ref{sec:complexCANDs}, after presenting the usual approach to algebraic classification based on the Jordan form of the Cotton--York tensor in Sec.~\ref{sec:CottonYorkJordan}.


\section{Cotton--York tensor}
\label{sec:CottonYork}

The number of independent components of the Cotton tensor $C_{abc}$ \eqref{Cotton_Definition} in three dimensions, being five, is exactly equal to the number of components of a symmetric and traceless rank-2 tensor. This can be obtained as the \emph{Hodge dual}.

More specifically, following the conventions given in \cite{Garcia:2017}, with only slight modifications, the \emph{Cotton--York tensor} (sometimes also called the Schouten--Cotton--York tensor) is geometrically defined by equation (20.111) in \cite{Garcia:2017} (but denoted as $C_{\alpha\beta}$ therein) as
\begin{equation} \label{Cotton-York_Definition}
Y_{ab}\equiv \bolde_a \rfloor\, ^*\mathbf{C}_b =\, ^*(\mathbf{C}_b \wedge \boldsymbol{\omega }_a) \, ,
\end{equation}
where ${\boldsymbol{\omega }_a = g_{ab} \, \boldsymbol{\omega }^b}$  is the linear combination of the basis 1-forms \eqref{basis} and $\mathbf{C}_b$ is the Cotton (``vector valued'') 2-form
\begin{equation} \label{Cotton_2-form}
\text{2-form:} \qquad \mathbf{C}_b \equiv \tfrac{1}{2}\,C_{m n b} \, \boldsymbol{\omega}^m \wedge \boldsymbol{\omega}^n \, .
\end{equation}
The symbol $\rfloor$ in \eqref{Cotton-York_Definition} stands for the \emph{interior product} defined on a general $p$-form $\boldsymbol{\sigma }$ as\footnote{The exterior calculus notation and definitions used here are mainly taken from Appendix A in \cite{Hehl}.}
\begin{equation} \label{Interior_product}
\bolde_a \rfloor \boldsymbol{\sigma } \equiv \frac{1}{(p-1)!} \, \sigma _{ab _{2} \ldots b _{p}} \, \boldsymbol{\omega }^{b _{2}} \wedge \ldots \wedge \boldsymbol{\omega }^{b _{p}} \, .
\end{equation}
Another common notation for this operation is ${\iota_{\footnotesize{\bolde}}\,\boldsymbol{\sigma}}$.

Recall that a metric-independent \emph{Hodge dual operator} can be defined by employing the so called $\epsilon$-basis. This is constructed by subsequent interior products of the Levi-Civita tensor
\begin{equation}
\boldsymbol{\omega} \equiv -3! \, \sqrt{-g} \,\, \boldsymbol{\omega }^0 \wedge \boldsymbol{\omega }^1 \wedge \boldsymbol{\omega }^2\, ,
\end{equation}
in which $g$ is  the determinant of the metric $g_{ab}$. In components, this tensor explicitly reads
\begin{equation}
\omega _{abc}=-\sqrt{-g} \,\, \varepsilon _{abc} \, , \qquad \text{or} \qquad
\omega ^{abc}= \frac{1}{\sqrt{-g}} \,\, \varepsilon ^{abc} \, ,
\end{equation}
where $\varepsilon ^{abc}=\varepsilon _{abc}$ is the Levi-Civita symbol. Without loss of generality we assume that the null triad has the following orientation
\begin{equation} \label{assum_(C)}
\omega _{abc} \,\,  k^a \, l^b \, m^c = 1 \, .
\end{equation}
It is equivalent to defining the Levi-Civita symbol in the null basis $\{ \boldsymbol{\omega }^{b } \}$ as $\varepsilon ^{123}=\varepsilon _{123}=-1$. Such an orientation ensures that, in an orthonormal frame, the spatial part will have the right-handed orientation. The correspondence between the Hodge dual operation and the $\epsilon$-basis is
\begin{alignat}{4}
\text{2-form:}& \qquad \boldsymbol{\epsilon }_{a} &&\equiv \, ^*\boldsymbol{\omega }_a  &&=\bolde_a \rfloor \boldsymbol{\omega} &&= \tfrac{1}{2}\omega _{a b c } \, \boldsymbol{\omega }^{b } \wedge \boldsymbol{\omega }^{c } \, , \nonumber\\
\text{1-form:}& \qquad \boldsymbol{\epsilon }_{ab} &&\equiv \, ^*(\boldsymbol{\omega }_a \wedge \boldsymbol{\omega }_b)&&=\bolde_b \rfloor \boldsymbol{\epsilon }_a &&= \omega _{ab c } \, \boldsymbol{\omega }^{c } \, , \\
\text{0-form:}& \qquad \epsilon _{abc} &&\equiv \, ^*(\boldsymbol{\omega }_a \wedge \boldsymbol{\omega }_b \wedge \boldsymbol{\omega }_c) &&=\bolde_c \rfloor \boldsymbol{\epsilon }_{ab} &&=\omega _{abc} \, .
\nonumber
\end{alignat}

Applying this construction of the Hodge dual to the (vector valued) Cotton 2-form \eqref{Cotton_2-form} we obtain $^* \mathbf{C}_b=\tfrac{1}{2}C_{m n b} \, \boldsymbol{\epsilon }^{m n} $, that gives the following expression for the dual
\begin{equation} \label{Dual_Cotton_1-form}
\text{1-form:}\qquad ^* \mathbf{C}_b = \tfrac{1}{2} \, \omega ^{mnk}  \, C_{mnb}\, g_{kc} \, \boldsymbol{\omega }^{c} \, .
\end{equation}
By performing the contractions ${\bolde_a \rfloor\,\boldsymbol{\omega }^{c} = {\delta_a}^c }$
we get from the definition \eqref{Cotton-York_Definition} an explicit prescription for the Cotton--York tensor, namely
\begin{align} \label{Cotton-York}
Y_{ab} &=\tfrac{1}{2} \, g_{ak}\, \omega^{kmn}  \, C_{mnb} \nonumber\\[1mm]
       &=-\sqrt{-g} \,\, \varepsilon_{amn} \,\Big( \nabla^m {R^n}_{b} -\frac{1}{4}\, {\delta^n}_{b}\,\partial^m R \Big) \, .
\end{align}
This alternative form of the Cotton tensor appeared in York's work \cite{York}, but was already discussed before by  Arnowitt, Deser and Misner \cite{ADM}. It encodes the same information as the Cotton tensor, but it is a rank-lower tensor. One of its major advantages is that it is \emph{symmetric}
\begin{equation} \label{York_symmetry}
Y_{ab}=Y_{ba} \, ,
\end{equation}
and also \emph{traceless}
\begin{equation}
{Y_{a}}^{a}=0 \, .
\end{equation}
Moreover, 2+1 spacetime is locally \emph{conformally flat} if and only if ${Y_{ab}=0}$. This Cotton--York tensor is the key tensor in the context of 2+1 gravity, whose algebraic classification has already been introduced and successfully employed. Since it is a rank-2 tensor, the eigenvalue problem can be formulated exactly as a standard eigenvalue problem for matrices, see \cite{GHHM, Garcia:2017}.

To find an explicit relation to our new method of classification, we first express $Y_{ab}$ in the null triad \eqref{null} as
\begin{equation} \label{York_null_basis}
Y_{ab}=\sum_{I ,J=1 }^3 Y_{I J} \,\, \omega^{I}_a \, \omega^{J}_b \, ,
\end{equation}
where $Y_{I J}$ are the corresponding components (recall that the dual basis is $\{ \boldsymbol{\omega }^{I}\}\equiv\{ -\boldl, -\boldk, \boldm \}$). We can uniquely relate them to the Newman--Penrose-like Cotton scalars \eqref{Psi}. Writing the sum \eqref{York_null_basis} explicitly, using just the symmetry property \eqref{York_symmetry} of the Cotton--York tensor, we get
\begin{align} \label{York_null_explicit}
Y_{ab}=&\ Y_{11} \, l_a \, l_b+Y_{12}(l_a \, k_b +k_a \, l_b)-Y_{13}(l_a \, m_b +m_a \, l_b) \nonumber \\
&+Y_{22} \, k_a \, k_b-Y_{23}(k_a \, m_b +m_a \, k_b)+Y_{33} \, m_a \, m_b \, .
\end{align}
Applying the normalization condition \eqref{null}, from \eqref{York_null_explicit} the coefficients can be expressed as\footnote{Basically, these are the real symmetric quantities $\Psi_{AB}$ introduced in Sec.~5 of \cite{Sousa:2008}.}
\begin{align}\label{Yab}
Y_{11}&=Y_{ab} \, k^a \, k^b \, , \nonumber\\
Y_{12}&=Y_{ab} \, k^a \, l^b \, , \nonumber\\
Y_{13}&=Y_{ab} \, k^a \, m^b \, , \nonumber\\
Y_{22}&=Y_{ab} \, l^a \, l^b \, , \\
Y_{23}&=Y_{ab} \, l^a \, m^b \, , \nonumber\\
Y_{33}&=Y_{ab} \, m^a \, m^b \, . \nonumber
\end{align}
Using the definition \eqref{Cotton-York} with the relation \eqref{assum_(C)} and the full expression \eqref{Cotton_General} of the general Cotton tensor, one arrives at the following result
\begin{align}
Y_{11}&=-\Psi _0 \, , & Y_{12}&=-\Psi _2 \, , & Y_{13}&=-\Psi _1 \, , \nonumber \\
Y_{22}&=\Psi _4 \, , & Y_{23}&=\Psi _3 \, , & Y_{33}&=-2 \, \Psi _2 \, .
\end{align}
The \emph{general Cotton--York tensor} \eqref{York_null_explicit} in the null triad basis thus takes the form
\begin{align} \label{York}
Y_{ab}=&-\Psi _0 \, l_a \, l_b +\Psi _1 (l_a \, m_b +m_a \, l_b)  \nonumber \\
& -\Psi _2 (l_a \, k_b + k_a \, l_b + 2 \, m_a \, m_b )\nonumber \\
& -\Psi _3 (k_a \, m_b + m_a \, k_b)+\Psi _4 \, k_a \, k_b \, .
\end{align}
This is the key expression that will allow us now to relate the Cotton scalars $\Psi_{\rm A}$ to the algebraic classification of spacetimes in 2+1 gravity, and to demonstrate its equivalence with the previous classification scheme based on the Jordan form of the Cotton--York tensor \cite{Garcia:2017}.

To this end, it is important to express the general Cotton--York tensor in an \emph{orthonormal basis} ${\{ \boldE_0, \, \boldE_1, \, \boldE_2 \}}$ corresponding to the null triad \eqref{null} via the usual relations
\begin{equation} \label{ortho}
\boldE_0\equiv\tfrac{1}{\sqrt{2}}(\boldk+\boldl) \, ,\qquad \boldE_1\equiv\tfrac{1}{\sqrt{2}}(\boldk-\boldl) \, , \qquad \boldE_2\equiv\boldm
\, .
\end{equation}
Due to the normalization \eqref{null}, such a basis satisfies the conditions
\begin{equation}
\boldE_0 \cdot \boldE_0 =-1 \, , \qquad \boldE_1 \cdot \boldE_1 =1 \, , \qquad \boldE_2 \cdot \boldE_2 =1 \, ,
\end{equation}
with all other scalar products equal to zero, i.e., the metric in this basis reads
\begin{equation}\label{ortho-metric}
g_{ab}= \text{diag}(-1,1,1) \, .
\end{equation}
It means that $\boldE_0$ is the (future-oriented) timelike unit vector, while $\boldE_1$ and $\boldE_2$ are perpendicular Cartesian spatial vectors. It also follows that
\begin{align}
E_{0}^{\,a} \, k_a&=-\tfrac{1}{\sqrt{2}} \, , & E_{0}^{\,a} \, l_a&=-\tfrac{1}{\sqrt{2}} \, , & E_{0}^{\,a} \, m_a&=0 \, , \nonumber\\
E_{1}^{\,a} \, k_a&= \tfrac{1}{\sqrt{2}} \, , & E_{1}^{\,a} \, l_a&=-\tfrac{1}{\sqrt{2}} \, , & E_{1}^{\,a} \, m_a&=0 \, , \\
E_{2}^{\,a} \, k_a&=0 \, , & E_{2}^{\,a} \, l_a&=0 \, , & E_{2}^{\,a} \, m_a&=1 \, . \nonumber
\end{align}
Using \eqref{York}, we thus easily obtain \emph{all orthonormal projections of the Cotton--York tensor}, such as ${Y_{00}\equiv E_{0}^{\,a} E_{0}^{\,b}\,Y_{ab} = - \Psi _2 -\tfrac{1}{2}(\Psi _0 - \Psi _4)}$, etc. The result is
\begin{equation}\label{Y_ab}
Y_{ab}=
\begin{pmatrix}
- \Psi _2 - \tfrac{1}{2}(\Psi _0 - \Psi _4)   & - \tfrac{1}{2}(\Psi _0 + \Psi _4) &  - \tfrac{1}{\sqrt{2}}(\Psi _1 - \Psi _3)\\[4mm]
- \tfrac{1}{2}(\Psi _0 + \Psi _4) & \Psi _2 - \tfrac{1}{2}(\Psi _0 - \Psi _4)  & -\tfrac{1}{\sqrt{2}}(\Psi _1 + \Psi _3)\\[4mm]
- \tfrac{1}{\sqrt{2}}(\Psi _1 - \Psi _3) & -\tfrac{1}{\sqrt{2}}(\Psi _1 + \Psi _3) & -2\Psi _2
\end{pmatrix} ,
\end{equation}
which is clearly a \emph{symmetric real matrix} (${Y_{ab}=Y_{ba}}$, ${{Y_a}^a=0}$). Actually, it is a direct 2+1 analogue of the symmetric complex ${3 \times 3}$ matrix~${\mbox{\boldmath$Q$}}$ (with zero trace) which is used for the algebraic classification of the Weyl tensor in 3+1 spacetimes in the original Petrov approach (see Eq.~(3.65) in \cite{Stephani}).


\section{Equivalence with the previous method of classification}
\label{sec:CottonYorkJordan}

To complete this work, it remains to prove the equivalence of our new convenient method of algebraic classification, based on the Cotton scalars~$\Psi_{\rm A}$ and the multiplicity of CANDs, with the previous  ``Petrov-type'' classification scheme based on finding the specific Jordan forms of the Cotton--York tensor. As summarized in Introduction, this was first considered in \cite{Barrow} and refined in \cite{GHHM}.

Let us repeat the main results, following sections 1.2.1 and 20.5.2  of the monograph \cite{Garcia:2017} by Garc\'ia-D\'iaz. The key idea is to solve the ordinary \emph{eigenvalue problem} ${{Y_a}^b\,v_b = \lambda\,v_a }$ for the Cotton--York ${3 \times 3}$~matrix ${{Y_a}^b \equiv Y_{ac} \, g^{cb}}$. In view of \eqref{Y_ab} and \eqref{ortho-metric}, in the orthonormal basis \eqref{ortho} we get its explicit expression ($a$ denotes rows, while $b$ denotes the columns)
\begin{equation}\label{class}
{Y_a}^b=
\begin{pmatrix}
\Psi _2 + \tfrac{1}{2}(\Psi _0 - \Psi _4)   & - \tfrac{1}{2}(\Psi _0 + \Psi _4) &  - \tfrac{1}{\sqrt{2}}(\Psi _1 - \Psi _3)\\[4mm]
\tfrac{1}{2}(\Psi _0 + \Psi _4) & \Psi _2 - \tfrac{1}{2}(\Psi _0 - \Psi _4)  & -\tfrac{1}{\sqrt{2}}(\Psi _1 + \Psi _3)\\[4mm]
\tfrac{1}{\sqrt{2}}(\Psi _1 - \Psi _3) & -\tfrac{1}{\sqrt{2}}(\Psi _1 + \Psi _3) & -2\Psi _2
\end{pmatrix} .
\end{equation}
It is important to emphasize that the matrix ${Y_a}^b$ is traceless but \emph{not symmetric}. Therefore, the roots of the characteristic \emph{cubic polynomial} ${\,\det \,({Y_a}^b - \lambda\,{\delta_a}^b ) = 0\,}$ may be \emph{complex}. Nevertheless, according to the possible eigenvalues $\lambda _1, \lambda _2$  and ${\lambda_3 = - \lambda _1 - \lambda _2}$, one can find the corresponding canonical \emph{Jordan forms}, defining the  algebraic ``Petrov'' types of all 2+1 geometries. Such forms are presented in Table~\ref{Tab:Petrov-Types}, which is actually the copy of Table~1.2.1 of \cite{Garcia:2017}.

\begin{table}[!h]
\begin{center}
\begin{tabular}{ccc}
\hline
\\[-8pt]
   ``Petrov'' type & Jordan form $J$ of ${Y_a}^b$ & eigenvalues relation   \\[2pt]
\hline
\\[-4pt]
   I
   & $\begin{pmatrix} \lambda _1 & 0 & 0\\ 0 & \lambda _2 & 0\\ 0 & 0 & -\lambda _1 -\lambda _2 \end{pmatrix}$
   & ${\quad\lambda _1 \neq \lambda _2, \,\quad  \lambda _3=-\lambda _1 - \lambda _2}$ \\[20pt]
   II
   & $\begin{pmatrix} \lambda _1 & 1 & 0\\ 0 & \lambda _1 & 0\\ 0 & 0 & -2\lambda _1 \end{pmatrix}$
   & ${\quad\lambda _1 = \lambda _2 \neq 0, \,\quad \lambda _3=-2\lambda _1 }$ \\[20pt]
   D
   & $\begin{pmatrix} \lambda _1 & 0 & 0\\ 0 & \lambda _1 & 0\\ 0 & 0 & -2\lambda _1 \end{pmatrix}$
   & ${\quad\lambda _1 = \lambda _2 \neq 0, \,\quad \lambda _3=-2\lambda _1}$ \\[20pt]
   III
   & $\begin{pmatrix} 0 & 1 & 0\\ 0 & 0 & 1\\ 0 & 0 & 0 \end{pmatrix}$
   & ${\lambda _1 = \lambda _2 =\lambda _3= 0}$ \\[20pt]
   N
   & $\begin{pmatrix} 0 & 1 & 0\\ 0 & 0 & 0\\ 0 & 0 & 0 \end{pmatrix}$
   & ${\lambda _1 = \lambda _2 =\lambda _3 = 0}$  \\[20pt]
   O
   & $\begin{pmatrix} 0 & 0 & 0\\ 0 & 0 & 0\\ 0 & 0 & 0 \end{pmatrix}$
   &  \\[20pt]
\hline
\end{tabular}
\caption{\label{Tab:Petrov-Types} Traditional algebraic classification of the Cotton--York tensor~${Y_a}^b$ based on the possible Jordan forms and eigenvalues.}
\end{center}
\end{table}


\begin{table}[!b]
\begin{center}
\begin{tabular}{ccc}
\hline
\\[-8pt]
   ``Petrov'' type & normal form $N$ of ${Y_a}^b$ & values of the Cotton scalars  \\[2pt]
\hline
\\[-4pt]
   I
   & $\begin{pmatrix} \lambda _1 & 0 & 0\\ 0 & \lambda _2 & 0\\ 0 & 0 & -\lambda _1 -\lambda _2 \end{pmatrix}$
   & $\begin{matrix} \Psi _1 =0=\Psi _3  \\ \Psi _0 =\frac{1}{2}(\lambda _1-\lambda _2)=-\Psi _4\\ \Psi _2=\frac{1}{2}(\lambda _1+\lambda _2) \end{matrix}$ \\[20pt]
   II
   & $\begin{pmatrix} \lambda _1 -1 & -1 & 0\\ 1 & \lambda _1 +1 & 0\\ 0 & 0 & -2\lambda _1 \end{pmatrix}$
   & $\begin{matrix} \Psi _0 = 0\,, \ \Psi _1 =0 = \Psi _3\\ \Psi _2=\lambda _1 \\ \Psi _4 =2 \end{matrix}$ \\[20pt]
   D
   & $\begin{pmatrix} \lambda _1 & 0 & 0\\ 0 & \lambda _1 & 0\\ 0 & 0 & -2\lambda _1 \end{pmatrix}$
   & $\begin{matrix} \Psi _0 =0=\Psi _4\\ \Psi _1 =0=\Psi _3 \\ \Psi _2=\lambda _1\end{matrix}$ \\[20pt]
   III
   & $\begin{pmatrix} 0 & 0 & 1\\ 0 & 0 & -1\\ -1 & -1 & 0 \end{pmatrix}$
   & $\begin{matrix} \Psi _0 =0=\Psi _4 \\ \Psi _1 =0=\Psi _2 \\ \Psi _3=\sqrt{2} \end{matrix}$ \\[20pt]
   N
   & $\begin{pmatrix} -1 & -1 & 0\\ 1 & 1 & 0\\ 0 & 0 & 0 \end{pmatrix}$
   & $\begin{matrix} \Psi _0 =0= \Psi _2 \\ \Psi _1 =0=\Psi _3\\ \Psi _4=2\end{matrix}$  \\[20pt]
   O
   & $\begin{pmatrix} 0 & 0 & 0\\ 0 & 0 & 0\\ 0 & 0 & 0 \end{pmatrix}$
   & all\  ${\Psi_{\rm A}=0}$  \\[20pt]
\hline
\end{tabular}
\caption{\label{Tab:Normal-Types} Algebraic classification based on the possible normal forms of the Cotton--York tensor~${Y_a}^b$ and the specific values of the Cotton scalars~$\Psi_{\rm A}$.}
\end{center}
\end{table}

Now, we would like to find a one-to-one correspondence between the canonical Jordan forms $J$ of ${Y_a}^b$ presented in Table~\ref{Tab:Petrov-Types} for each ``Petrov''  type, and the canonical values of the Cotton scalars~$\Psi_{\rm A}$. By comparing the Jordan form of type~I with the explicit expression \eqref{class} we uniquely obtain the conditions ${\Psi _1 \pm \Psi _3 =0}$ (so that ${\Psi _1 = 0 = \Psi _3}$), ${\Psi _0 + \Psi _4 = 0}$, ${\tfrac{1}{2}(\Psi _0 - \Psi _4) + \Psi _2 = \lambda_1}$ and ${-\tfrac{1}{2}(\Psi _0 - \Psi _4) + \Psi _2 = \lambda_2}$ (so that ${2\Psi _2=\lambda _1+\lambda _2}$ and ${\Psi _0-\Psi _4=\lambda _1-\lambda _2}$). Similarly for type~D we immediately obtain ${\Psi _0 =0=\Psi _4}$, ${\Psi _1 =0=\Psi _3}$ and ${\Psi _2=\lambda _1}$. However, for types~II,~III and~N such an identification is \emph{not directly possible}. Instead, in these cases it is necessary to employ an equivalent (alternative) \emph{normal forms of the Cotton--York matrix}~${Y_a}^b$.

More precisely, we look for a \emph{similarity transformation} between the Jordan form~$J$ and the specific \emph{normal form}~$N$, such that
\begin{equation}
N = A \, J \, A^{-1} \, ,
\end{equation}
where $A$ is an invertable matrix and $A^{-1}$ its inverse. In particular, a direct calculation shows that for type~II geometries such a similarity transformation takes the form
\begin{equation}
N=
\begin{pmatrix} \lambda_1 -1 & -1 & 0\\ 1 & \lambda_1 +1 & 0\\ 0 & 0 & -2\lambda _1 \end{pmatrix} =
\begin{pmatrix} -1 & 1 & 0\\ 1 & 0 & 0\\ 0 & 0 & 1 \end{pmatrix}
\begin{pmatrix} \lambda_1 & 1 & 0\\ 0 & \lambda _1 & 0\\ 0 & 0 & -2\lambda _1 \end{pmatrix}
\begin{pmatrix} 0 & 1 & 0\\ 1 & 1 & 0\\ 0 & 0 & 1 \end{pmatrix}.
\end{equation}
The subcase ${\lambda _1=0}$ gives the transformation for type~N geometries. And for type~III we get
\begin{equation}
N=
\begin{pmatrix} 0 & 0 & 1\\ 0 & 0 & -1\\ -1 & -1 & 0 \end{pmatrix} =
\begin{pmatrix} -1 & 0 & 1\\ 1 & 0 & 0\\ 0 & -1 & 0 \end{pmatrix}
\begin{pmatrix} 0 & 1 & 0\\ 0 & 0 & 1\\ 0 & 0 & 0 \end{pmatrix}
\begin{pmatrix} 0 & 1 & 0\\ 0 & 0 & -1\\ 1 & 1 & 0 \end{pmatrix}.
\end{equation}
Such normal forms of the Cotton--York tensor~${Y_a}^b$ can be uniquely identified with the canonical values of the Cotton scalars~$\Psi_{\rm A}$ for each algebraic type. The results are summarized in Table~\ref{Tab:Normal-Types}.

\newpage

We have thus proven that for each ``Petrov'' algebraic type in ${D=3}$ there exists a privileged orthonormal basis, which can be called the \emph{principal Cotton--York basis}, in which ${Y_a}^b$ has the corresponding normal form $N$, and the associated canonical values of the Cotton scalars~$\Psi_{\rm A}$, as given in the last column of Table~\ref{Tab:Normal-Types}. Actually, it is an analogue of Table~4.2 in \cite{Stephani} which contains the normal forms of the Weyl tensor for all Petrov types in ${D=4}$.

Moreover, the specific values of the Cotton scalars~$\Psi_{\rm A}$ in Table~\ref{Tab:Normal-Types} are \emph{fully consistent} with our new simpler method of algebraic classification of 2+1 geometries, as summarized in Table~\ref{Tab-classification} and corroborated in Table~\ref{Tab:algebraic-types} to also include the related multiplicity of the Cotton-aligned null directions CANDs. To be more specific:

\begin{itemize}

\item {\bf Type I} geometries with CAND $\boldk'$ are defined by the existence of the principle null triad such that~${\Psi'_0=0}$, see Sec.~\ref{CAND}. By inspecting the last column of Table~\ref{Tab:Normal-Types} we observe that such a condition is not satisfied in the principal Cotton--York basis because~${\Psi_0\ne0}$ (recall that ${\lambda _1 \neq \lambda _2}$ for type~I spacetimes). However, we can employ a suitable Lorentz transformation~\eqref{eq:null-rotatin-fixed-l} which for the canonical values of the Cotton scalars reduces to
\begin{align}
\Psi_0' &= (K^4+1)\, \Psi_0 +6K^2\,\Psi_2  \, , \nonumber\\
\Psi_1' &= \sqrt{2}\,K\, ( K^2\,\Psi_0 + 3\,\Psi_2) \, , \nonumber\\
\Psi_2' &= K^2\,\Psi_0 + \Psi_2 \, ,\\
\Psi_3' &= -\sqrt{2}\,K\,\Psi_0 \, , \nonumber\\
\Psi_4' &= -\Psi_0 \, . \nonumber
\end{align}
Obviously, we achieve~${\Psi'_0=0}$ by taking $K$ to be any root of the bi-quadratic equation
\begin{align}
 \Psi_0\,K^4 +6\Psi_2\,K^2 + \Psi_0 = 0 \, .
\end{align}
Because ${\Psi _0 =\frac{1}{2}(\lambda _1-\lambda _2)}$ and ${\Psi _2=\frac{1}{2}(\lambda _1+\lambda _2)}$, these \emph{four distinct} explicit roots are
\begin{align}
 K^2 = \frac{-3\Psi_2 \pm \sqrt{9\Psi_2^2- \Psi_0^2 }}{\Psi_0}
     = -3\,\frac{\lambda _1+\lambda _2}{\lambda _1-\lambda _2} \pm \frac{ \sqrt{9(\lambda _1+\lambda _2)^2- (\lambda _1-\lambda _2)^2 }}{\lambda _1-\lambda _2} \, .
\end{align}
The corresponding null vectors ${\boldk' = \boldk + \sqrt2\, K\,\boldm + K^2\, \boldl }$ are then CANDs because~${\Psi'_0=0}$. Moreover, ${K^2\,\Psi_0 + 3\,\Psi_2 = \pm \tfrac{1}{2} \sqrt{9(\lambda _1+\lambda _2)^2- (\lambda _1-\lambda _2)^2 }}$, so that generally~${\Psi'_1\ne0}$.
\vspace{2mm}

\item {\bf Type II} geometries in the principal Cotton--York basis have ${\Psi _0 = \Psi _1 = 0}$, ${\Psi _2=\lambda _1 \ne 0}$, see Table~\ref{Tab:Normal-Types}. This fully corresponds to our definition presented in Table~\ref{Tab-classification}. In fact, we can even achieve ${\Psi _3=0}$ by performing (the inverse of) the Lorentz transformation~\eqref{eq:null-rotatin-fixed-l}, namely ${\Psi_3 = \Psi_3' - \sqrt{2}\,K\,\Psi_4'}$ for the particular choice of the null rotation parameter ${\sqrt{2}\,K = \Psi_3'/\Psi_4'}$.
    Using \eqref{eq:null-rotatin-fixed-l}, the condition~${\Psi'_0=0}$ for CAND $\boldk'$ in the principal Cotton--York basis reduces to a special form
\begin{align}
 (K^2-3\lambda_1)\,K^2 =0 \, .
\end{align}
    The factor $K^2$ shows that $\boldk$ is a \emph{double CAND}, and other two distinct CANDs are obtained by applying the null rotation with the parameters ${K=\pm \sqrt{3\lambda _1}}$.
\vspace{2mm}

\item {\bf Type D} geometries in the principal Cotton--York basis have ${\Psi _0=0=\Psi _1}$, ${\Psi _3=0=\Psi _4}$ and
${\Psi _2=\lambda _1 \ne 0}$. This is a complete agreement with our definition presented in Table~\ref{Tab-classification}.
There are \emph{two distinct} CANDs~$\boldk$ and~$\boldl$, both of multiplicity $2$ because ${K^2=0}$ and also ${L^2=0}$, see relations~\eqref{Class_eq-special-D} and~\eqref{Class_eq-special-D-l}.
\vspace{2mm}

\item {\bf Type III} geometries in the principal Cotton--York basis have ${\Psi _0 = \Psi _1 = \Psi_2 = 0}$, ${\Psi _3=\sqrt{2}}$ and ${\Psi_4=0}$, which exactly corresponds to our definition in Table~\ref{Tab-classification}. We can achieve ${\Psi _4=0}$ by performing (the inverse of) the Lorentz transformation~\eqref{eq:null-rotatin-fixed-k}. Indeed, for the particular choice of the null rotation parameter ${2\sqrt{2}\,L = \Psi_4'/\Psi_3'}$ we get ${\Psi_4 = \Psi_4' -2\sqrt{2}\,L\,\Psi_3' =0}$. The key equation~${\Psi'_0=0}$ given by \eqref{eq:null-rotatin-fixed-l} reduces to
\begin{align}
 K^3 =0 \, .
\end{align}
It demonstrates that $\boldk$ is a \emph{triple CAND}, while the fourth distinct CAND is $\boldl$ corresponding to ${L=0}$.
\vspace{2mm}

\item {\bf Type N} geometries in the principal Cotton--York basis have ${\Psi _0 = \Psi _1 = \Psi_2 = \Psi_3= 0}$ and ${\Psi _4=2}$, in full agreement with Table~\ref{Tab-classification}. We can achieve the fixed canonical value ${\Psi _4=2}$ from any ${\Psi _4'\ne0}$ by the boost \eqref{boost}, which implies a simple rescaling ${\Psi_4 = B^{2}\,\Psi'_4 }$, see \eqref{eq:boost-of-Psi}. In this case the condition~${\Psi'_0=0}$ reduces to
\begin{align}
 K^4 =0 \, ,
\end{align}
which proves that $\boldk$ is a \emph{quadruple CAND}.

\end{itemize}

These results are summarized in Table~\ref{Tab:Multiplicity}. They show that our new simpler method of algebraic classification of 2+1 geometries, based on the direct evaluation \eqref{Psi} of the Cotton scalars~$\Psi_{\rm A}$ and using the conditions in Table~\ref{Tab-classification}, is fully equivalent to the previous (rather cumbersome) ``Petrov'' approach based on determining the eigenvalues and the respective Jordan form $J$ of the Cotton--York tensor~${Y_a}^b$ (employed, e.g., in \cite{Garcia:2017}). Moreover, our approach shows the unique relation of the algebraic types to the corresponding multiplicity of the Cotton-aligned null directions (CANDs), in a complete analogy with the multiplicities of the principal null directions (PNDs) in ${D=4}$ gravity (see Section~4.3 of \cite{Stephani}) and the  Weyl-aligned null directions (WANDs) in ${D>4}$ gravity theories (see \cite{OrtaggioPravdaPravdova:2013}).

\begin{table}[h]
\begin{center}
\begin{tabular}{ccccc}
\hline
\\[-8pt]
   type & condition~${\Psi'_0=0}$ & roots & CANDs & multiplicity\\[2pt]
\hline
\\[-8pt]
I & $K^4+6b\,K^2+1=0$ & $K=\pm \sqrt{-3b \pm 2\sqrt{D}}$, &
\hbox{
   \rotatebox[origin=c]{-30}{$\leftarrow$}\hspace{-3mm}
   \raisebox{1.5mm}{\rotatebox[origin=c]{-60}{$\leftarrow$}}\hspace{-1mm}
   \raisebox{1.5mm}{\rotatebox[origin=c]{60}{$\rightarrow$}}\hspace{-3mm}
   \rotatebox[origin=c]{30}{$\rightarrow$}} &
1+1+1+1\\[-2mm]
\\
& & where $b=\frac{\lambda _1+\lambda _2}{\lambda _1-\lambda _2}$ and &\\[-3mm]
\\
& & $\sqrt{D}=\frac{\sqrt{2\lambda _1^2+5\lambda _1\lambda _2+2\lambda _2^2}}{\lambda_1-\lambda _2}$ &
\\[8pt]

II & $(K^2-3\lambda _1)\,K^2=0$ & \quad ${K=\pm \sqrt{3\lambda _1}}$ and double ${K=0}$ &
\hbox{
   \rotatebox[origin=c]{-30}{$\leftarrow$}\hspace{-3mm}
   \raisebox{1.5mm}{\rotatebox[origin=c]{-60}{$\leftarrow$}}\hspace{-1mm}
   \raisebox{0.6mm}{\rotatebox[origin=c]{45}{$\Rightarrow$}}} &
1+1+2
\\[8pt]

D & ${L^2=0}$ and ${K^2=0}$ & \quad double ${L=0}$ and double ${K=0}$  &
\hbox{
   \rotatebox[origin=c]{-45}{$\Leftarrow$}\,\rotatebox[origin=c]{45}{$\Rightarrow$}} &
2+2
\\[8pt]

III & ${L=0}$ and $K^3=0$ &  ${L=0}$ and triple $K=0$ &
\hbox{
   \rotatebox[origin=c]{-30}{$\leftarrow$}\hspace{-1mm}
   \raisebox{0.6mm}{\rotatebox[origin=c]{45}{$\Rrightarrow$}}} &
1+3
\\[8pt]

N & $K^4=0$ & quadruple ${K=0}$ &
{\Large
   \hbox{
   \rotatebox[origin=c]{45}{$\Rightarrow$}\hspace{-6.1mm}
   \raisebox{-0.30mm}{\rotatebox[origin=c]{45}{$\Rightarrow$}}}
   } &
4
\\[2pt]
\hline
\end{tabular}
\caption{Multiplicities of the Cotton-aligned null directions (CANDs) for all algebraic types. These are obtained from the key condition ${\Psi'_0=0}$ expressed in the principal Cotton--York basis, in which the Cotton scalars~$\Psi_{\rm A}$ have the canonical form presented in Table~\ref{Tab:Normal-Types}. Multiplicity of the root corresponds to the multiplicity of the related CAND. If the root is ${K=0}$ then the null vector~$\boldk$ is CAND. Similarly, the root ${L=0}$ identifies that $\boldl$ is CAND. For type N geometries the only nontrivial Cotton scalar is ${\Psi _4}$, so that the vector~$\boldk$ is a quadruple CAND corresponding to the multiplicity 4 of the root ${K=0}$.}
\label{Tab:Multiplicity}
\end{center}
\end{table}

\newpage


\section{Invariants assisting with the algebraic classification}
\label{invariants}

To complete our new procedure of algebraic classification of 2+1 geometries, we now investigate an important concept of \emph{scalar curvature polynomial invariants} which can be constructed from the Cotton tensor and the related Cotton--York tensor. In fact, it will turn out that these invariants play a crucial role in easily determining the algebraic type of the spacetime.

From the explicit expression \eqref{Cotton_General} for the Cotton tensor, using the normalization relations \eqref{null-comp} and \eqref{Bivector_Contraction}, we can directly evaluate the quadratic scalar invariant
\begin{align}
C_{abc}\,C^{abc} = 4\,( \Psi _0 \Psi _4 -2\,\Psi _1 \Psi _3 -3\,\Psi _2^2)
\, ,\label{Invariant-Cotton_General}
\end{align}
and similarly from the expression \eqref{York} for the Cotton--York tensor we similarly obtain
\begin{align}
Y_{ab}\,Y^{ab} = -2\,( \Psi _0 \Psi _4 -2\,\Psi _1 \Psi _3 -3\,\Psi _2^2)
\, .\label{Invariant-Cotton-York_General}
\end{align}
Another invariant can be constructed as their specific cubic combination
\begin{align}
C_{abc} \,C^{abd}\, {Y^c}_{d} = 6\,(\Psi _0\Psi _3^2 -\Psi _1^2\Psi _4 +2\,\Psi _0\Psi _2\Psi _4 +2\,\Psi _1\Psi _2\Psi _3 +2\,\Psi _2^3) \, . \label{Invariant-Cotton-York-Cotton_General}
\end{align}

It can be immediately seen that for \emph{type N spacetimes}, in which the only non-vanishing Cotton scalar is $\Psi_4$,  one gets
\begin{align}
C_{abc}\,C^{abc} = 0 = Y_{ab}\,Y^{ab}    \qquad \hbox{and}\qquad   C_{abc} \,C^{abd}\, {Y^c}_{d} =0\, .
\end{align}
In fact, \emph{all algebraic types} can be uniquely identified by using such invariants expressed in terms of the specific polynomials constructed from the Cotton scalars~$\Psi_{\rm A}$. The two key invariants are
\begin{align}
I &= \tfrac{1}{4}\, C_{abc}\, C^{abc} = -\tfrac{1}{2}\,Y_{ab}\, Y^{ab} \, ,\nonumber\\
J &= \tfrac{1}{6}\, C_{abc}\, C^{abd}\, {Y^c}_d \, .
\end{align}
In view of \eqref{Invariant-Cotton_General}, \eqref{Invariant-Cotton-York_General} and \eqref{Invariant-Cotton-York-Cotton_General} they can be defined as
\begin{align}
I & \equiv \Psi _0\Psi _4 -2\,\Psi _1\Psi _3 -3\,\Psi _2^2 \, ,\nonumber\\
J & \equiv 2\,\Psi _0\Psi _2\Psi _4 +2\,\Psi _1\Psi _2\Psi _3 +2\,\Psi _2^3 + \Psi _0\Psi _3^2 -\Psi _4\Psi _1^2   \, . \label{Invariants_I_J}
\end{align}

These unique invariants naturally occur in the expression for the \emph{discriminant $\Delta $ of the key quartic equation} \eqref{Class_eq}, after dropping the primes. Indeed, a direct calculation shows that
\begin{align}\label{discriminant}
-\Delta = 2^8\,  I^3 + 2^6 3^3\, J^2 \, .
\end{align}
It is well-known that the necessary and sufficient condition for the quartic equation to have a \emph{multiple root} is ${\Delta =0}$. In the present context it means that a spacetime is \emph{algebraically special} (it admits at least one \emph{multiple CAND}), i.e. it is at least of type~II, if and only if
\begin{align} \label{I3=J2}
4\, I^3 =-27\, J^2 \, .
\end{align}
Furthermore, if and only if ${I=0=J}$ the key equation \eqref{Class_eq} has at least a \emph{triple root} and the corresponding spacetime is of type~III or of type~N. To distinguish them we define additional quantities
\begin{align}
G &\equiv \Psi _1\Psi _4^2 -3\,\Psi _2\Psi _3\Psi _4 -\Psi _3^3 \, ,\nonumber\\
H &\equiv 2\,\Psi _2\Psi _4 + \Psi _3^2 \, , \label{Invariants_G_H_N}\\
N &\equiv 3\, H^2 + \Psi _4^2\, I \, .\nonumber
\end{align}
A spacetime is of algebraic type N if ${I=0=J}$ and ${G=0=H}$. Finally, algebraically special spacetime with ${I\neq 0 \neq J}$ is of type~D if and only if ${G=0=N}$ (otherwise it remains of type~II).

These conditions follow from first eliminating the \emph{cubic term} in the quartic equation \eqref{Class_eq}, resulting in the so called \emph{depressed quartic}. The quantity $G$ is the coefficient in front of the \emph{linear term} in this depressed quartic. Its vanishing reduces the equation to bi-quadratic equation, from which the subsequent analysis depending on ${H=0}$ and ${N=0}$ immediately follows.

\begin{figure}[t!]
\includegraphics[scale=0.95]{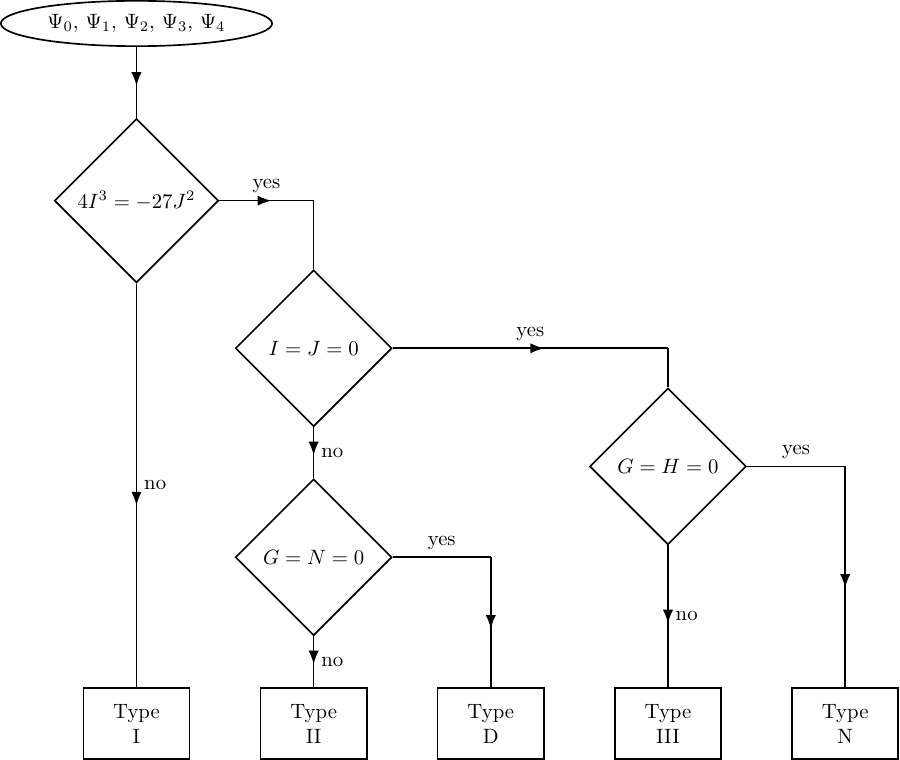}
\caption{Flow diagram for determining the algebraic type of a 2+1 geometry using the invariants~\eqref{Invariants_I_J} and \eqref{Invariants_G_H_N} constructed from the Cotton scalars~$\Psi_{\rm A}$. In the special case ${\Psi_4 =0 =\Psi_0}$ it is necessary to employ Table~\ref{Tab:Algorithm-special-case}.}
\label{Flow_diagram}
\end{figure}

The useful complete\footnote{The procedure is not applicable if ${\Psi _4=0}$ and ${\Psi _0 \neq 0}$. However, in such a case it is possible to perform the swap ${\Psi _0 \leftrightarrow -\Psi _4}$ and ${\Psi _1 \leftrightarrow -\Psi _3}$ in the expressions, after which the algorithm in Fig. \ref{Flow_diagram} can be used.} algorithm of algebraic classification is synoptically summarized by the flow diagram in Fig.~\ref{Flow_diagram}. Actually, it is a one-to-one analogue of the flow diagram for the algebraic classification of ${D=4}$ spacetimes presented in the original work \cite{InvernoRussel:1971} by d'Inverno and Russell-Clark, and in Fig.~9.1 of \cite{Stephani}.

With the help of the practical algorithm in Fig.~\ref{Flow_diagram}, we can finally confirm the classification into the algebraic types contained in Table~\ref{Tab:Normal-Types}. The invariants $I, J, G, H, N$ for the corresponding special values of~$\Psi_{\rm A}$ in the principal Cotton--York basis (contained in the last column of Table~\ref{Tab:Normal-Types}) are shown in the last column of Table~\ref{Tab:Normal-Types-Invariants}. Their values and mutual relations are fully consistent with the flow diagram scheme in Fig.~\ref{Flow_diagram}.

\begin{table}[!h]
\begin{center}
\begin{tabular}{ccc}
\hline
\\[-8pt]
   algebraic type & special values of ${\Psi_{\rm A}}$ & corresponding invariants \\[2pt]
\hline
\\[-4pt]
   I
   & $\begin{matrix} \Psi _1 =0=\Psi _3  \\ \Psi _0 =\frac{1}{2}(\lambda _1-\lambda _2)=-\Psi _4\\ \Psi _2=\frac{1}{2}(\lambda _1+\lambda _2) \end{matrix}$
   & $\begin{matrix} \quad I=\lambda _1\lambda _2 - (\lambda_1 + \lambda_2)^2  \\ \quad J=\lambda _1\lambda _2\,(\lambda _1+\lambda _2) \end{matrix}$
   \\[20pt]
   II
   & $\begin{matrix} \Psi _0 = 0\,, \ \Psi _1 =0 = \Psi _3\\ \Psi _2=\lambda _1 \\ \Psi _4 =2 \end{matrix}$
   & $\begin{matrix} I=-3\,\lambda_1^2  \\ J=2\,\lambda_1^3 \\  G=0\,,\ N = 36\,\lambda_1^2 \end{matrix}$
   \\[26pt]
   D
   & $\begin{matrix} \Psi _0 =0=\Psi _4\\ \Psi _1 =0=\Psi _3 \\ \Psi _2=\lambda _1\end{matrix}$
   & $\begin{matrix} I=-3\,\lambda_1^2  \\ J=2\,\lambda_1^3 \\  G=0=N \end{matrix}$
   \\[20pt]
   III
   & $\begin{matrix} \Psi _0 =0=\Psi _4 \\ \Psi _1 =0=\Psi _2 \\ \Psi _3=\sqrt{2} \end{matrix}$
   & $\begin{matrix} I=0=J \\ G=-2\sqrt{2}\,,\  H=2 \end{matrix}$
   \\[20pt]
   N
   & $\begin{matrix} \Psi _0 =0= \Psi _2 \\ \Psi _1 =0=\Psi _3\\ \Psi _4=2\end{matrix}$
   & $\begin{matrix} I=0=J \\ G=0=H\\ \end{matrix}$
   \\[20pt]
\hline
\end{tabular}
\caption{\label{Tab:Normal-Types-Invariants} Consistency of the algebraic classification based on the normal forms of the Cotton--York tensor~${Y_a}^b$ with the invariants calculated from the corresponding specific values of the Cotton scalars~$\Psi_{\rm A}$.}
\end{center}
\end{table}


\section{Complex CANDs and complex eigenvalues}
\label{sec:complexCANDs}

We have already mentioned at the end of Sec.~\ref{CAND-multiplicity} that in 2+1 gravity the algebraic classification suffers from the (somewhat unwelcome) property that the key \emph{real} equation \eqref{Class_eq} can, in general, have some \emph{complex roots}~$K$. More specifically, an equation of the fourth order can have either four (possibly multiple) real roots, or four complex roots, or two real and two complex roots. This implies that some of the null vectors ${\boldk = \boldk' - \sqrt2\, K\,\boldm' + K^2\, \boldl'}$, representing the Cotton-aligned principal null directions CANDs obtained by \eqref{tetradtrans-repeated}, \emph{may be complex}. This cannot happen neither in 3+1 gravity (because the Newman--Penrose formalism in ${D=4}$ with the Weyl scalars $\Psi_{\rm A}$ is complex, and thus complex roots $K$ are allowed, leading to four real PNDs) nor in higher dimensional gravity ${D>4}$ (because the corresponding real quartic equation involves more parameters $K_i$, leading to four real WANDs --- except in type~G spacetimes).

It is thus natural to suggest a \emph{subclassification} which, for each algebraic type, \emph{distinguishes the real and complex CANDs}. In particular, we may introduce the definition:

\begin{itemize}
\item subtypes~I$_{\rm r}$, II$_{\rm r}$ and D$_{\rm r}$: all four (possibly multiple) \emph{CANDs are real},

\item subtypes~I$_{\rm c}$, II$_{\rm c}$ and D$_{\rm c}$: some of the \emph{CANDs are (formally) complex}.
\end{itemize}

We need not distinguish such subtypes for geometries of algebraic type~III and type~N. Indeed, if a quartic equation \eqref{Class_eq} admits a root of multiplicity three or four, necessarily all roots must be real, and thus {III\,$\equiv$\,III$_{\rm r}$} and {N\,$\equiv$\,N$_{\rm r}$}.

On the other hand, for type~I and type~II (and thus also for type~D) geometries we have to investigate the complexity of the roots of the equation \eqref{Class_eq}, which is best done using the invariants \eqref{Invariants_I_J} and \eqref{Invariants_G_H_N}.\footnote{We assume that $\Psi _4 \neq 0$, otherwise we perform the swap $\Psi _0 \leftrightarrow -\Psi _4$ and $\Psi _1 \leftrightarrow -\Psi _3$, provided $\Psi _0 \neq 0$.}

\begin{itemize}
\item For \emph{type~I} (when ${4I^3\neq -27 J^2}$) the equation \eqref{Class_eq} has four distinct roots. If
\begin{equation} \label{Class_eq-2-complex_roots}
4I^3 > -27J^2 \, ,
\end{equation}
the discriminant $\Delta$ given by \eqref{discriminant} is negative, and the equation has two distinct real roots and a pair of (conjugated) complex roots. It means that the geometry is of subtype~I$_{\rm c}$ with two complex CANDs.

If the relation \eqref{Class_eq-2-complex_roots} is \emph{not} satisfied, there exist either four distinct real roots, or four distinct complex roots. If both the relations
\begin{equation} \label{Class_eq-4-real_roots}
H>0 \qquad \text{and} \qquad N>0
\end{equation}
hold then there are four real roots and the geometry is of subtype~I$_{\rm r}$ with four real CANDs. Otherwise, all roots of the equation \eqref{Class_eq} are complex and the algebraic subtype is $\text{I}_{\rm c}$.

\item For \emph{type~II} (when ${4I^3 = -27 J^2}$) the discriminant $\Delta$ vanishes and there is at least one double root. The only possibility of the subtype~II$_{\rm c}$ is when there is one double real root and a pair of complex conjugated roots and thus CANDs. This happens if and only if
\begin{equation} \label{Class_eq-special-II-subtype}
N < 0 \, .
\end{equation}
For ${G=0=N}$ the geometry is of type~D (see Fig.~\ref{Flow_diagram}). The subtype~D$_{\rm c}$ with two double complex CANDs occurs when the equation \eqref{Class_eq} admits two conjugated complex roots of multiplicity two. This can only happen in the case when
\begin{equation}
H < 0 \, .
\end{equation}

In particular, with the canonical form of the Cotton scalars ${\Psi_0=0=\Psi_1}$ with ${\Psi_2\ne0}$, the subtype II$_{\rm r}$ occurs if and only if ${3\Psi_2\Psi_4+\Psi_3^2 >0}$.  In the opposite case ${3\Psi_2\Psi_4+\Psi_3^2 <0}$ it is of subtype~II$_{\rm c}$.
The case ${3\Psi_2\Psi_4+\Psi_3^2 =0}$  gives the geometry of subtype~D$_{\rm r}$ with two double real CANDs.
\end{itemize}

\vspace{2mm}

Actually, the presence of the complex CANDs in the subtypes I$_{\rm c}$, II$_{\rm c}$ and D$_{\rm c}$ is (indirectly) related to the known property that the Cotton--York matrix ${Y_a}^b$ \emph{is not symmetric} and thus its \emph{eigenvalues can be complex}, see \cite{GHHM} and Sec.~20.5.2 in the monograph \cite{Garcia:2017}. Therefore, in these works a special algebraic subtype denoted as Class~$\text{I}^{\prime }$ was introduced. This represents the case when the Cotton--York matrix has three distinct eigenvalues --- of which one is real and two are complex (necessarily complex conjugated).

In view of the Jordan  form $J$ of ${Y_a}^b$ (equal to its normal form $N$) and the corresponding Cotton scalars $\Psi_{\rm A}$, given in Tables~\ref{Tab:Petrov-Types} and~\ref{Tab:Normal-Types}, we can relate the \emph{complex eigenvalues} of Class~$\text{I}^{\prime }$, written as
\begin{equation} \label{Cotton-I'-eigenvalues}
\lambda_1\equiv \lambda_r + \complex\,\lambda_c
\qquad \hbox{and} \qquad
\lambda_2\equiv \lambda_r - \complex\,\lambda_c \, ,
\end{equation}
(so that ${\lambda_3=-\lambda_1-\lambda_2 = -2 \lambda_r}$) to the canonical values of the Cotton scalars as ${\Psi_1=0=\Psi_3}$,
\begin{equation} \label{Cotton-I'}
\Psi_2=\lambda_r\, , \qquad
\Psi_0=\complex\,\lambda_c\, , \qquad
\Psi_4=-\complex\,\lambda_c\, .
\end{equation}
It means that the Cotton scalars $\Psi_0$ and $\Psi_4$ \emph{are purely imaginary} (and complex conjugated).

However, in this case it is more appropriate to employ the equivalent form of ${Y_a}^b$ given in table in Sec.~20.5.2 in \cite{Garcia:2017}, namely
\begin{equation}
{Y_a}^b =
\begin{pmatrix}
\lambda _r & \lambda _c & 0\\
-\lambda _c & \lambda _r & 0\\
0 & 0 & -2\lambda _r
\end{pmatrix}
 .
\end{equation}
In view of \eqref{class}, the corresponding \emph{real} Cotton scalars take the values ${\Psi _1 =0 =\Psi _3}$ and
\begin{equation} \label{Cotton-I'real}
\Psi_2=\lambda_r\, , \qquad
\Psi_0= \lambda_c =\Psi _4 \, .
\end{equation}
In this case the key equation \eqref{Class_eq} for determining the CANDs becomes
\begin{equation} \label{Class_eqI'}
\lambda_c\,K^4  - 6\lambda_r\,K^2 - \lambda_c =0 \, ,
\end{equation}
and the four solutions to this bi-quadratic equation are
\begin{align}
 K_\pm^2 = 3\frac{\lambda _r}{\lambda _c} \pm \sqrt{9\frac{\lambda _r^2}{\lambda _c^2}+1} \, ,
\end{align}
so that ${K_+^2 > 0}$ and ${K_-^2 < 0}$. There are thus \emph{two complex CANDs corresponding to}~${\pm \complex\,|K_-|}$. This shows that the Class~I$^{\prime}$ defined in \cite{GHHM, Garcia:2017} is equivalent to the case ${4I^3 > -27J^2}$ of subtype~I$_{\rm c}$, introduced here. If (and only if) ${\lambda_c=0}$ then the only non-trivial (real) Cotton scalar \eqref{Cotton-I'} or \eqref{Cotton-I'real} is ${\Psi_2=\lambda_r}$. The eigenvalues are ${\lambda_1 = \lambda_2 = \Psi_2}$ and ${\lambda_3= -2 \Psi_2}$. They are real, and the spacetime is of type~D.

Moreover, it can be seen that our subtypes I$_{\rm r}$ and I$_{\rm c}$ are directly related to the Petrov--Segre types I$_{\mathbb R}$ and I$_{\mathbb C}$ in TMG, as introduced in \cite{ChowPopeSezgin:2010a, ChowPopeSezgin:2010b}, with real and complex eigenvalues of the Cotton--York/traceless Ricci tensors, respectively.

On the other hand, it follows from Table~\ref{Tab:Normal-Types} that type~II and type~D spacetimes have only real eigenvalues and real Cotton scalars~$\Psi_{\rm A}$, so \emph{it is not necessary} to introduce analogous Class~$\text{II}^{\prime }$ and Class~$\text{D}^{\prime }$. However, the CANDs given by the complex roots of \eqref{Class_eq} \emph{can be complex}. It thus seems that it is useful to define the subtypes~I$_{\rm c}$, II$_{\rm c}$, D$_{\rm c}$, respectively, to denote these subcases.



\section{Explicit examples of the new classification method}
\label{sec:example}

Finally, to demonstrate the usefulness of our simple classification scheme based on the Cotton scalars~$\Psi_{\rm A}$, as summarized in  Table~\ref{Tab-classification} (see also Table~\ref{Tab:algebraic-types}) and in the flow diagram Fig.~\ref{Flow_diagram}, we will apply it to several explicit classes of 2+1 geometries.

\subsection{Robinson--Trautman spacetimes with $\Lambda$ and electromagnetic field}

Let us consider a large class of  the Robinson--Trautman spacetimes with a cosmological constant~$\Lambda$ and an \emph{aligned} electromagnetic field. Recently in \cite{Podolsky&Papajcik} we derived that in the geometrically adapted canonical coordinates ${r, u, x}$ the \emph{most general form} of such 2+1 solutions to Einstein--Maxwell equations (with a coupling constant ${\kappa_0>0}$) can be written as
\begin{align}
\dd s^2 =&\  \frac{r^2}{P^2}\,
\big( \dd x + e \,P^2 \dd u \big)^2 -2\,\dd u\,\dd r \nonumber \\
& +\Big(\mu\, Q^2 - \kappa_0\,Q^2 \ln \Big|\frac{Q}{r}\Big|
  +2\,(\ln Q)_{,u}\,r + \Lambda\,r^2\Big)\, \dd u^2\, ,  \label{RTmetric-final}
\end{align}
with the Maxwell field potential
\begin{equation}
 \bA =  Q \,\ln\frac{r}{r_0}\,  \dd u \,,
\label{RTformA-final}
\end{equation}
so that ${\,\bF = (Q/r)\, \dd r \wedge \dd u\,}$, see Eqs.~(180) and (182) of \cite{Podolsky&Papajcik}.
Here $\mu$ is a constant, $Q(u)$ is \emph{any} function of $u$, and the metric functions $P(u,x)$, $e(u,x)$ satisfy the field equation
\begin{equation}\label{RTFconstraint-2Q}
\Big(\frac{Q}{P}\Big)_{,u} = Q\,(e\,P)_{,x} \, .
\end{equation}

Now, using the general components of the Ricci tensor $R_{ab}$ (see Eqs.~(A24)--(A29) of \cite{Podolsky&Papajcik}) the Cotton tensor $C_{abc}$ corresponding to the solution \eqref{RTmetric-final} can be calculated from the definition \eqref{Cotton_Definition}. Its non-vanishing coordinate components are
\begin{align}
C_{urr} & = \frac{\kappa _0 Q^2}{2 \, r^3} \, , \nonumber\\
C_{uru} & = L_r  \, \frac{\kappa _0 Q^4}{2 \, r^3} + A_{uru} \, \frac{1}{r^2}
+ (e^2P^2-\Lambda )\, \frac{\kappa _0 Q^2}{2 \, r} \, , \nonumber\\
C_{xru} & = e \,\frac{\kappa _0 Q^2}{2\, r} \, , \nonumber\\
C_{xuu} & = A_{xuu} - \Big( P(eP)_{,x} + \frac{P_{,u}}{P} \Big) _{,x} \,\Big( \frac{\kappa _0}{2}  + L_r \Big)  \frac{Q^2}{r}  \, ,\\
C_{urx} & = e \, \frac{\kappa _0 Q^2}{2 \, r} \, ,\nonumber\\
C_{xrx} & = \frac{\kappa _0 Q^2}{2 \, r\, P^2 } \, , \nonumber\\
C_{xux} & = \Big( (eP)_{,x} + \frac{P_{,u}}{P^2} \Big)\Big( \frac{3}{2} \, \kappa _0 + L_r \Big) \frac{Q^2}{P} - (2 \, \kappa _0 +L_r) \frac{Q \, Q_{,u}}{P^2} \, , \nonumber
\end{align}
where the function $L_r(r)$ is
\begin{equation}
L_r \equiv \kappa _0 \, \ln \Big| \frac{Q}{r} \Big| - \mu \, ,
\end{equation}
and the more involved functions $A_{uru}$ and $A_{uux}$ are
\begin{align}
A_{uru} &\equiv (\kappa _0 + L_r) Q \, Q_{,u} - \Big( P(eP)_{,x}+\frac{P_{,u}}{P} \Big)\Big( \frac{3}{2} \, \kappa _0 + L_r \Big)  Q^2 \, , \nonumber\\
A_{xuu} &\equiv \Big( P(eP)_{,x} + \frac{P_{,u}}{P} \Big)\Big( \frac{3}{2} \, \kappa _0 + L_r \Big) \, e\,Q^2
- (2\kappa _0 + L_r)\,e \, Q \, Q_{,u} \\
& \, \quad + \Big[ P(eP)_{,x} + \frac{P_{,u}}{P} \Big] _{,x} \, \frac{Q_{,u}}{Q} + P_{,x} \, (P \, e_{,u})_{,x} + P \, (P_{,x} \, e_{,u})_{,x} + P \, (P \, e_{,ux})_{,x} \nonumber \\
& \, \quad +\Big(\frac{P_{,uu}}{P} \Big)_{,x}-4\frac{P_{,u}}{P} \Big(\frac{P_{,u}}{P} \Big)_{,x} - \Big[2e^2P^2P_{,x}^2 + P^4(e_{,x}^2 + e\,e_{,xx})+e \, P^3(5e_{,x}P_{,x}+e \, P_{,xx})\Big]_{,x}  \, .\nonumber
\end{align}

These coordinate expressions of the Cotton tensor components are very complicated. However, using the \emph{natural null triad} satisfying \eqref{null},
\begin{equation}
\boldk=\partial_r\,, \qquad
\boldl=\frac{1}{2}\,g_{uu}\,\partial_r+\partial_u \,, \qquad \boldm=\frac{1}{\sqrt{g_{xx}}}\,\big(g_{ux}\,\partial_r+\partial_x\big) \, ,
\end{equation}
(see Eq.~(6) of \cite{Podolsky&Papajcik}), the definition \eqref{Psi} and the field equation \eqref{RTFconstraint-2Q}, we obtain \emph{simple Cotton scalars}~$\Psi_{\rm A}$, namely
\begin{align}\label{RT-EM-Psi}
\Psi _0& = 0\, , \nonumber\\
\Psi _1& = -\frac{\kappa _0 Q^2}{2r^3} \, ,\nonumber\\
\Psi _2& = -eP \,\frac{\kappa _0 Q^2}{2r^2} \, , \\
\Psi _3& = \Big( \kappa _0\ln \Big| \frac{Q}{r}\Big|  - \mu \Big)\,\frac{\kappa _0 Q^4}{4r^3}
+\big(e^2P^2  - \Lambda\big)\,\frac{\kappa _0Q^2}{4r} \, , \nonumber\\
\Psi _4& = eP \,\Big( \kappa _0\ln \Big| \frac{Q}{r} \Big| - \mu \Big)\,\frac{\kappa _0 Q^4}{2r^2}
+ eP\, \big(e^2P^2 - \Lambda \big)\,\frac{\kappa _0 Q^2}{2} \, . \nonumber
\end{align}

Because ${\Psi _0=0}$, it is obvious from Table~\ref{Tab-classification} that \emph{all} such Robinson--Trautman spacetimes with a cosmological constant~$\Lambda$ and an aligned electromagnetic field are (at least) of \emph{algebraic type~I}. Moreover, it follows from Sections~\ref{CAND} and~\ref{CAND-multiplicity} that the null vector ${\boldk=\partial_r}$ is CAND. In other words, this Cotton-aligned null direction ${\boldk=\partial_r}$ coincides with the privileged null-aligned direction of the electromagnetic field.

In fact, the general Cotton scalars \eqref{RT-EM-Psi} \emph{can considerably be further simplified} just by performing a suitable Lorentz transformation of the triad. In particular, the null rotation \eqref{kfixed} with ${\boldk}$ fixed, changing ${\boldl}$ and ${\boldm}$ as
\begin{align}
\boldk' = \boldk \, , \qquad
\boldl' = \boldl + \sqrt2\, L\,\boldm + L^2\, \boldk  \, ,\qquad\,
\boldm' = \boldm + \sqrt2\, L\,\boldk \, , \label{kfixedagain}
\end{align}
transforms the Cotton scalars according to the rule \eqref{eq:null-rotatin-fixed-k}. Choosing the specific real parameter~$L$,
\begin{align}
 \sqrt{2}\,L = - eP\,r \, , \label{choice-of-L}
\end{align}
and relabeling the constant $\mu$ to the function ${m(u)\equiv \mu\, Q^2(u)}$, we get a very nice result
\begin{align}\label{eq:null-rotatin-fixed-k-RT-EM}
\Psi_0' &= 0 \, , \nonumber\\[1mm]
\Psi_1' &= -\frac{\kappa_0 Q^2}{2r^3}  \, , \nonumber\\[2mm]
\Psi_2' &= 0 \, ,\\[1mm]
\Psi_3' &= - \Big( \,m -\kappa_0 Q^2 \ln \Big| \frac{Q}{r} \Big| +\Lambda\,r^2 \, \Big)\,\frac{\kappa_0 Q^2}{4r^3}
\, , \nonumber\\[2mm]
\Psi_4' &= 0 \,\, . \nonumber
\end{align}
Interestingly,
\begin{align}\label{eq:Psi3*Psi1}
2\,\Psi_3' = \Big( \,m -\kappa_0 Q^2 \ln \Big| \frac{Q}{r} \Big| +\Lambda\,r^2 \, \Big)\,\Psi_1'\,.
\end{align}
These are the Cotton scalars expressed with respect to the unique null triad
\begin{align}
\boldk' & = \partial_r \, , \nonumber\\
\boldl' & = \partial_u
  + \frac{1}{2}\,\Big( \,m - \kappa_0\,Q^2 \ln \Big|\frac{Q}{r}\Big| + 2\,(\ln Q)_{,u}\,r + \Lambda\,r^2  \Big)\,\partial_r
  - eP^2\,\partial_x
\, ,\label{unique-triad}\\
\boldm' & = \frac{P}{r}\,\partial_x \, . \nonumber
\end{align}

Clearly, all these scalars vanish when ${Q=0}$, which corresponds to \emph{vacuum} solutions with $\Lambda$, and thus necessarily are spacetimes of constant curvature (locally Minkowski, de~Sitter or anti-de~Sitter), which are \emph{conformally flat}.

In the non-trivial case ${Q\ne0}$ with an (aligned) electromagnetic field, the key scalar curvature invariants \eqref{Invariants_I_J} are
\begin{align}
I &=  -2\,\Psi_1'\Psi_3' =
   \Big( - m +\kappa _0 Q^2 \ln \Big| \frac{Q}{r} \Big| -\Lambda\,r^2 \, \Big)\,
  \frac{\kappa_0^2 Q^4}{4r^6}\, , \nonumber\\ 
J &=  0  \, . \label{Invariant_J-RT}
\end{align}
It is obvious that the fundamental condition \eqref{I3=J2}, that is ${\,4\, I^3 =-27\, J^2}$, \emph{cannot be satisfied}. Consequently, \emph{all} such spacetimes are of algebraic type~I, see also the flow diagram in Fig.~\ref{Flow_diagram}. More precisely, here it is necessary to employ Table~\ref{Tab:Algorithm-special-case} because in this case ${\Psi_4' =0 =\Psi_0'}$. Using the fact that ${\Psi_1'\ne0}$, ${\Psi_2'=0}$, ${\Psi_3'\ne0}$, the corresponding row in Table~\ref{Tab:Algorithm-special-case} determines the type~I.

In our work \cite{Podolsky&Papajcik} we were able to identify the famous class of (cyclic symmetric) \emph{charged black hole electrostatic solutions}~\cite{Peldan:1993} that is the 2+1 analogue to the Reissner--Nordstr\"om--(anti-)de~Sitter solution, see the metric~(192) in \cite{Podolsky&Papajcik} and the review given in Sec.~11.2 of \cite{Garcia:2017}. It arises as the special subcase ${Q=\hbox{const.}}$ and ${e=0}$ of the metric \eqref{RTmetric-final}. In such a situation $L$ given by \eqref{choice-of-L} is trivial (${L=0}$), and \eqref{RT-EM-Psi} is thus identical to \eqref{eq:null-rotatin-fixed-k-RT-EM}. In any case, the key invariants \eqref{Invariant_J-RT} remain the same, which implies that these electrostatic black hole spacetimes are of algebraic type~I. The same result was obtained already by applying the Petrov classification based on the corresponding Jordan form of the Cotton--York tensor, see Sec.~11.1.5 in \cite{Garcia:2017}.

Interestingly, \emph{on the horizons} which are localized by the condition ${\,-m +\kappa _0 Q^2 \ln \big| \frac{Q}{r} \big| - \Lambda\,r^2=0}$ the scalar ${\Psi_3'}$ vanishes, see \eqref{eq:Psi3*Psi1}, so that according to Table~\ref{Tab:Algorithm-special-case} these horizons are of algebraic type~III.
Moreover, for \eqref{eq:null-rotatin-fixed-k-RT-EM} the key equation \eqref{Class_eq} determining the CANDs becomes
\begin{equation}
\Big[ \Big( - m +\kappa _0 Q^2 \ln \Big| \frac{Q}{r} \Big| -\Lambda\,r^2 \, \Big)\,K^2 + 2\, \Big]\, K=0\, .
\end{equation}
The square bracket tells us that \emph{above the horizon}, where ${-m +\kappa _0 Q^2 \ln \big| \frac{Q}{r} \big| - \Lambda\,r^2>0}$, there exist \emph{two complex CANDs}, so that such a region of the spacetime is of algebraic subtype~I$_{\rm c}$ (which in this case is equivalent to Class~I$^\prime$). Contrarily, below the horizon there are four real CANDs, and therefore the region is of subtype I$_{\rm r}$.

\subsection{Other examples of 2+1 spacetimes of various algebraic types}

In their seminal work \cite{GHHM}, Garc\'ia, Hehl, Heinicke and Mac\'ias investigated some  solutions of Einstein's field equations in 2+1 gravity, as well as solutions of the topologically massive gravity (TMG) model of Deser, Jackiw and Templeton, presenting explicit examples for each algebraic class. To further confirm and justify our classification method, we will now apply it to these examples studied in Sec.~7 of \cite{GHHM}.

\subsubsection{Type~I (and type~D) spacetime}

The line element given by Eqs.~(114)--(117) in \cite{GHHM} takes the form
\begin{align}\label{metric-I}
\dd s^2 =& -(a_1+a_2)^2\dd \psi ^2 -2(a_1+a_2)^2\sinh \theta \, \dd \psi \, \dd \phi \nonumber \\
& -(a_1^2-a_2^2)\sin 2\psi \, \cosh \theta \, \dd \theta \, \dd \phi
  +(a_1^2 \sin ^2\psi +a_2^2\cos ^2\psi )\, \dd \theta ^2 \\
&+\big[(a_1^2\cos ^2\psi +a_2^2\sin ^2\psi )\cosh ^2 \theta -(a_1+a_2)^2\sinh ^2\theta \big]\, \dd \phi^2 \, , \nonumber
\end{align}
with the natural orthonormal dual basis
\begin{align}
\boldsymbol{\omega }^0 &= -(a_1+a_2)(\dd \psi +\sinh \theta \, \dd \phi )\, ,\nonumber \\
\boldsymbol{\omega }^1 &= a_1 (-\sin \psi \, \dd \theta +\cos \psi \, \cosh \theta \, \dd \phi )\, ,\\
\boldsymbol{\omega }^2 &= a_2 (\cos \psi \, \dd \theta +\sin \psi \, \cosh \theta \, \dd \phi )\, .\nonumber
\end{align}
In view of \eqref{ortho}, the corresponding null triad reads
\begin{align}
\boldk &= \frac{1}{\sqrt{2}}\Big[ \Big( \frac{1}{a_1+a_2}-\frac{\cos \psi \, \tanh \theta }{a_1}\Big) \partial _{\psi }-\frac{1}{a_1}\sin \psi \, \partial _{\theta }+\frac{1}{a_1}\cos \psi \, \sech \theta \, \partial _{\phi }\Big]\, ,\nonumber \\
\boldl &= \frac{1}{\sqrt{2}}\Big[ \Big( \frac{1}{a_1+a_2}+\frac{\cos \psi \, \tanh \theta }{a_1}\Big) \partial _{\psi }+\frac{1}{a_1}\sin \psi \, \partial _{\theta }-\frac{1}{a_1}\cos \psi \, \sech \theta \, \partial _{\phi }\Big ]\, , \\
\boldm &= -\frac{1}{a_2}\sin \psi \, \tanh \theta \, \partial _{\psi } +\frac{1}{a_2}\cos \psi \, \partial _{\theta } +\frac{1}{a_2}\sin \psi \, \sech \theta \, \partial _{\phi }\, . \nonumber
\end{align}
The non-vanishing Cotton tensor components of the metric \eqref{metric-I} are
\begin{align}
C_{\psi \theta \theta } &= 2\, \frac{a_1^3-a_2^3}{a_1 a_2(a_1+a_2)}\, \sin 2\psi \, ,\nonumber\\
C_{\psi \phi \theta } &= 2\, \frac{(a_1^2+a_1a_2+a_2^2)[a_1+a_2-(a_1-a_2)\cos 2\psi \,]}{a_1a_2(a_1+a_2)}\cosh \theta \, ,\nonumber\\
C_{\psi \theta \phi } &= -2\,\frac{(a_1^2+a_1a_2+a_2^2)[a_1+a_2+(a_1-a_2)\cos 2\psi \,]}{a_1a_2(a_1+a_2)}\cosh \theta \, ,\nonumber\\
C_{\psi \phi \phi } &= -2\,\frac{a_1^3-a_2^3}{a_1 a_2(a_1+a_2)}\sin 2\psi \, \cosh ^2 \theta \, ,\\
C_{\theta \phi \psi } &= 4\,\frac{a_1^2+a_1 a_2 +a_2^2}{a_1 a_2} \cosh \theta \, ,\nonumber\\
C_{\theta \phi \theta } &= -2\,\frac{a_1^3-a_2^3}{a_1 a_2(a_1+a_2)}\sin 2\psi \, \sinh \theta \, ,\nonumber\\
C_{\theta \phi \phi } &= \frac{(a_1^2+a_1 a_2 +a_2^2)[3(a_1+a_2)+(a_1-a_2)\cos 2\psi \,]}{a_1\, a_2(a_1+a_2)}\sinh 2\theta \, .\nonumber
\end{align}
The Cotton scalars \eqref{Psi} evaluated with respect to this basis are simply the constants
\begin{align}
\Psi _0 &= -2\,\frac{a_1^3+(a_1+a_2)^3}{a_1^2 a_2^2(a_1+a_2)^2} \, ,\nonumber\\
\Psi _1 &= 0 \, ,\nonumber\\
\Psi _2 &= -2\,\frac{a_1^2+a_1 a_2+a_2^2}{a_1^2 a_2 (a_1+a_2)^2} \, ,\\
\Psi _3 &= 0 \, ,\nonumber\\[3mm]
\Psi _4 &= -\Psi _0 \, ,\nonumber
\end{align}
and the scalar invariants \eqref{Invariants_I_J} reduce to
\begin{align}\label{invariants-typeI}
I = -16 \,\frac{(a_1^2+a_1 a_2 +a_2^2)^3}{a_1^4 a_2^4 (a_1+a_2)^4}\, , \qquad
J =  64 \,\frac{(a_1^2+a_1 a_2 +a_2^2)^3}{a_1^5 a_2^5(a_1+a_2)^5} \, .
\end{align}
It is straightforward to check that the key relation ${4 I^3=-27 J^2}$ cannot be satisfied in general, so according to the flow diagram in Fig.~\ref{Flow_diagram} the spacetime \eqref{metric-I} is of \emph{algebraic type~I}.

We can obtain the four distinct CANDs by performing the null rotation \eqref{lfixed} to achieve~${\Psi'_0=0}$. Because ${\Psi_1=0=\Psi_3}$ and ${\Psi_4=-\Psi_0}$, the parameter $K$ has to satisfy the bi-quadratic equation
\begin{equation}
\Psi _0 +6\Psi _2 K^2 +\Psi _0 K^4 =0 \, ,
\end{equation}
see \eqref{eq:null-rotatin-fixed-l}. Its four distinct roots are $\pm K$ such that the two distinct values of $K^2$ are
\begin{equation} \label{Sol_K}
K^2 = \frac{1}{B}\,\bigg({-\frac{3}{a_1}-\frac{3}{a_2}-\frac{3a_2}{a_1^2}\pm
2\,\frac{a_1^2+a_1 a_2 +a_2^2}{a_1^2 a_2^2}\,\sqrt{2a_2^2-a_1 a_2 -a_1^2}}\,\bigg) \, ,
\end{equation}
where
\begin{equation}
B = \frac{3}{a_1}+\frac{3}{a_2}+\frac{2a_1}{a_2^2}+\frac{a_2}{a_1^2} \, .
\end{equation}

Actually, the metric \eqref{metric-I} is an example of a subtype~I$_{\rm c}$ spacetime that is \emph{not} equivalent to Class~I$^\prime $. It is straightforward to check that the relation \eqref{Class_eq-2-complex_roots} does not hold. The conditions \eqref{Class_eq-4-real_roots} are satisfied if and only if
\begin{equation}
a_1 < 0 \qquad \text{and} \quad -\frac{a_1}{2}<a_2<-2a_1\, ,
\end{equation}
or
\begin{equation}
a_1 > 0 \qquad \text{and} \quad -2a_1<a_2<-\frac{a_1}{2}\, ,
\end{equation}
and the spacetime is of subtype~I$_{\rm r}$. Otherwise, it is of subtype~I$_{\rm c}$ but not of Class I$^\prime$.

Interestingly, in the \emph{special case} ${a_1=a_2}$ the invariants \eqref{invariants-typeI} reduce to
\begin{align}\label{invariants-typeI-D}
I = -\frac{27}{a_1^6}\, , \qquad
J =  \frac{54}{a_1^9} \, ,
\end{align}
so that ${4 I^3=-27 J^2}$. According to the flow diagram in Fig.~\ref{Flow_diagram}, the spacetime \eqref{metric-I} becomes \emph{algebraically special}. In fact, because ${I, J \ne 0}$ and ${G=N=0}$, it degenerates to \emph{type~D}. This is in full agreement with the results of  \cite{GHHM}. Actually, the roots \eqref{Sol_K} are ${K=\pm {\rm i}}$ so that the two double CANDs are \emph{complex}, and the spacetime is of a subtype~D$_{\rm c}$ (see Sec.~\ref{sec:complexCANDs}).

From \eqref{Sol_K} we also conclude that the spacetime \eqref{metric-I} is of \emph{algebraic type~D if and only if}
${2a_2^2-a_1 a_2 -a_1^2=0}$. This has only two solutions, namely ${a_1=a_2}$ (discussed above) and ${a_1=-2a_2}$. In the latter case we get
\begin{align}\label{invariants-typeI-Db}
I = -\frac{27}{a_2^6}\, , \qquad
J =  \frac{54}{a_2^9} \, ,
\end{align}
again implying ${4 I^3=-27 J^2}$.

\subsubsection{Type~I$^\prime$ spacetime}

A generic example of the Class~I$^\prime$ metric is a spherically symmetric spacetime
\begin{equation} \label{metric-I'}
\dd s^2 = -\psi (r)\, \dd t^2 + \frac{\dd r^2}{\psi (r)} +r^2 \, \dd \varphi ^2 \, ,
\end{equation}
see Eq.~(111) in \cite{GHHM}, which suggests a natural orthonormal dual basis (beware of the opposite signature used in \cite{GHHM})
\begin{equation}
\boldsymbol{\omega }^0 = \sqrt{\psi }\, \dd t \, , \qquad \boldsymbol{\omega }^1 = \frac{\dd r}{\sqrt{\psi }} \, ,\qquad \boldsymbol{\omega }^2 = r\, \dd \varphi \, .
\end{equation}
The corresponding null triad reads
\begin{equation}
\boldk = \frac{1}{\sqrt{2}}\Big( \frac{1}{\sqrt{\psi}}\, \partial_t - \sqrt{\psi}\, \partial_r \Big)\, ,\qquad
\boldl = \frac{1}{\sqrt{2}}\Big( \frac{1}{\sqrt{\psi}}\, \partial_t + \sqrt{\psi}\, \partial_r \Big)\, ,\qquad
\boldm = \frac{1}{r}\, \partial _{\varphi }\, .
\end{equation}
The only non-vanishing components of the Cotton tensor are
\begin{equation}
C_{rtt} = \tfrac{1}{4}\psi \, \psi ^{\prime \prime \prime }  \, ,\qquad
C_{r\varphi \varphi} = \tfrac{1}{4}r^2\, \psi ^{\prime \prime \prime }\, ,
\end{equation}
where prime denotes the derivative with respect to~$r$. The Cotton scalars \eqref{Psi} simply evaluate to
\begin{align}
\Psi _0 &= 0\, ,\nonumber \\
\Psi _1 &= -\tfrac{1}{4\sqrt{2}}\, \sqrt{\psi}\, \psi ^{\prime \prime \prime }\, ,\nonumber \\
\Psi _2 &= 0\, ,\\
\Psi _3 &= -\Psi _1 \, ,\nonumber \\
\Psi _4 &= 0\, . \nonumber
\end{align}
According to Table~\ref{Tab:Algorithm-special-case} the spacetime is of type~I. In view of the equation \eqref{Class_eq} for CANDs which reduces to
\begin{equation}
(K^2+1)\, K = 0 \, ,
\end{equation}
we conclude that there are two real CANDs, namely $\boldk$ and $\boldl$, and two complex CANDs given by the solutions $K=\pm \complex$. Therefore, the spacetime \eqref{metric-I'} is of algebraic subtype~I$_{\rm c}$, which in this case is also equivalent to the Class~I$^\prime$.

\subsubsection{Type~I spacetime with type~II hypersurface}

The metric
\begin{equation}\label{metric-II}
\dd s^2 = -\e^{-4y}\, \dd t^2 -2\e^{-2y}\, \dd t\, \dd x +(\e^{2y}-1)\, \dd x^2 +\dd y^2 \, ,
\end{equation}
given by the orthonormal dual basis
\begin{align}
\boldsymbol{\omega }^0 = \e ^{-2y}\, \dd t +\dd x\, ,\qquad \boldsymbol{\omega }^1 = \e ^y\, \dd x\, ,\qquad \boldsymbol{\omega }^2 = \dd y \, ,
\end{align}
see Eq.~(132) in \cite{GHHM}, has the natural null triad
\begin{align} \label{Null_TypeII}
\boldk = \tfrac{1}{\sqrt{2}}\big[\e ^y(\e ^y-1)\, \partial _t +\e ^{-y}\, \partial _x \big]\, ,\quad
\boldl = \tfrac{1}{\sqrt{2}}\big[\e ^y(\e ^y+1)\, \partial _t -\e ^{-y}\, \partial _x \big]\, ,\quad
\boldm = \partial _y \, .
\end{align}
The non-vanishing components of the Cotton tensor are
\begin{align}
C_{txy} &= 4\e ^{-4y}(1-3\e ^{2y})\, ,\nonumber\\
C_{tyt} &= 6\e ^{-6y}(3-\e ^{2y})\, ,\nonumber\\
C_{tyx} &= 2\e ^{-4y}(7-9\e ^{2y})\, ,\\
C_{xyt} &= 2\e ^{-4y}(5-3\e ^{2y})\, ,\nonumber\\
C_{xyx} &= 6\e ^{-2y}(1-\e ^{4y})\, . \nonumber
\end{align}
The corresponding Coton scalars \eqref{Psi} read
\begin{align}
\Psi _0 &= -6\e ^{-3y}(1-3\e ^y +\e ^{2y} +\e ^{3y})\, ,\nonumber\\
\Psi _1 &= 0\, ,\nonumber\\
\Psi _2 &= \e ^{-3y}(6\e ^{2y}-2)\, ,\\
\Psi _3 &= 0\, ,\nonumber\\
\Psi _4 &= 6\e ^{-3y}(1+3\e ^y +\e ^{2y} -\e ^{3y})\, , \nonumber
\end{align}
and the scalar invariants \eqref{Invariants_I_J} are
\begin{align}
I &= 12\,(3-4e ^{-6y} +27\e ^{-4y} -30\e ^{-2y})\, , \nonumber\\
J &= 16\,\e ^{-9y}(8 -81\e ^{2y} +225 \e ^{4y} -171\e ^{6y} +27\e ^{2y})\, .
\end{align}
The key expression ${4I^3+27J^2 = 432^2\,\e ^{-14y}(1+\e ^{2y})^4(\e ^{6y} -7\e ^{4y} +7\e ^{2y} -1)}$ is nonzero, which implies that the spacetime \eqref{metric-II} is \emph{generally of algebraic type~I}.

However, on the special hypersurface ${y=0}$ this expression reduces to ${4I^3+27J^2 = 0}$.
In fact, the only non-vanishing Cotton scalars on ${y=0}$ are ${\Psi_2 =4}$ and ${\Psi_4 =24}$, and
thus ${I=-48}$, ${J=128}$, ${G=0}$, ${N=288^2}$. According to the flow diagram in Fig.~\ref{Flow_diagram} the spacetime \eqref{metric-II} is of \emph{type~II on} ${y=0}$. Since the condition \eqref{Class_eq-special-II-subtype} is not satisfied, it is of subtype~II$_{\rm r}$. Actually, the null vector $\boldk$ of the null triad \eqref{Null_TypeII} is the Cotton aligned null direction (CAND) on this hypersurface because ${\Psi_0=0}$ there.

\subsubsection{Type~I spacetime with type~III hypersurface}

Let us assume a simple metric
\begin{equation}\label{typeI-typeIII-metric}
\dd s^2 = -(t-x)^2\, \dd t^2 +(t+x)^2\, \dd x^2 +\dd y^2 \, ,
\end{equation}
given by the orthonormal basis (133) in \cite{GHHM}, namely
\begin{equation}
\boldsymbol{\omega }^0 = (x-t)\, \dd t\, ,\qquad \boldsymbol{\omega }^1 = (x+t)\, \dd x\, ,\qquad \boldsymbol{\omega }^2 = \dd y\, .
\end{equation}
The corresponding null triad reads
\begin{align}
\boldk = \frac{1}{\sqrt{2}}\Big( \frac{1}{t-x}\, \partial _t -\frac{1}{t+x}\, \partial _x \Big) \, ,\qquad
\boldl = \frac{1}{\sqrt{2}}\Big( \frac{1}{t-x}\, \partial _t +\frac{1}{t+x}\, \partial _x \Big) \, ,\qquad
\boldm = \partial _y \, .
\end{align}
The non-vanishing components of the Cotton tensor are
\begin{align}
C_{txt} &= -\frac{4(2t^2x+x^3)}{(t+x)^4(t-x)^2}\, ,\nonumber\\
C_{txx} &= \frac{4(t^3+2tx^2)}{(t+x)^2(t-x)^4}\, ,\nonumber\\
C_{tyy} &= -\frac{4(t^3+2tx^2)}{(t+x)^4(t-x)^4}\, ,\\
C_{xyy} &= \frac{4(2t^2x+x^3)}{(t+x)^4(t-x)^4}\, ,\nonumber
\end{align}
and thus the Cotton scalars \eqref{Psi}  take the form
\begin{align}
\Psi _0 &= 0\, ,\nonumber\\
\Psi _1 &= -2\sqrt{2}\,\, \frac{t^4+3t^3x+3tx^3-x^4}{(t+x)^5(t-x)^5}\, ,\nonumber\\
\Psi _2 &= 0\, ,\label{TypeIII_Psi1}\\
\Psi _3 &= -2\sqrt{2}\,\, \frac{t^4-t^3x+4t^2x^2+tx^3+x^4}{(t+x)^5(t-x)^5}\, ,\nonumber\\
\Psi _4 &= 0\, .\nonumber
\end{align}
In this case we have to be careful because ${\Psi _0 =0= \Psi _4}$ and the algorithm presented in Fig.~\ref{Flow_diagram} is thus not applicable. In such an exceptional case we have to use the Table~\ref{Tab:Algorithm-special-case} instead, which implies that the spacetime \eqref{typeI-typeIII-metric} is generally of \emph{type~I}.

However, applying this table we may notice that there exist specific subcases which are algebraically special. For example, the condition ${\Psi_1=0}$ with ${\Psi_3\ne0}$ leads to the algebraic \emph{type~III}. In view of \eqref{TypeIII_Psi1}, this can be achieved by taking the constraint ${x=t\,(\sqrt{13}+3)/2}$, which is exactly the conclusion presented in \cite{GHHM}. Indeed, \emph{on this hypersurface} the Cotton scalars reduce to
\begin{equation}
\Psi _0 = 0\, ,\qquad \Psi _1 = 0\, ,\qquad \Psi _2 = 0\, ,\qquad \Psi _3 = \frac{64\, \sqrt{2}\, (11+3\sqrt{13})}{27(3+\sqrt{13})^5\, t^6}\, ,\qquad \Psi _4 = 0 \, .
\end{equation}

\subsubsection{Type~D spacetime}

Taking the orthonormal basis given by Eqs.~(119)--(121) in \cite{GHHM} (with an opposite signature), we obtain the 2+1 G\"odel metric
\begin{align}\label{Godel}
\dd s^2  =&
-\frac{9}{\mu ^2} \, \dd t^2
+\frac{36}{\mu ^2} \big( \sqrt{r^2+1}-1\big) \, \dd t\, \dd \phi \\
&+\frac{9}{\mu ^2} \big( 8\sqrt{r^2+1}-3r^2-8\big) \, \dd \phi ^2
+\frac{9}{\mu ^2}\,\frac{\dd r^2}{r^2+1} \, .\nonumber
\end{align}
The null triad constructed from the natural basis, namely
\begin{align}
\boldsymbol{\omega }^0 = \frac{3}{\mu }\, \Big[ \dd t -2\big(\sqrt{r^2+1}-1\big) \dd \phi \Big] \, ,\qquad
\boldsymbol{\omega }^1 = \frac{3}{\mu }\, \frac{\dd r}{\sqrt{r^2+1}}\, ,\qquad
\boldsymbol{\omega }^2 = \frac{3}{\mu }\, r\, \dd \phi \, ,
\end{align}
has the form
\begin{align}
\boldk &= \frac{1}{\sqrt{2}}\,\frac{\mu}{3}\, \Big[ \partial_t -\sqrt{r^2+1}\, \partial_r\Big] \, ,\nonumber \\
\boldl &= \frac{1}{\sqrt{2}}\,\frac{\mu}{3}\, \Big[ \partial_t +\sqrt{r^2+1}\, \partial_r\Big] \, ,\\
\boldm &= \frac{\mu }{3r}\, \Big[ 2\big(\sqrt{r^2+1}-1\big)\, \partial_t +\partial_\phi \Big] \, . \nonumber
\end{align}
The non-vanishing components of the Cotton tensor are
\begin{align}
C_{t\phi r} &= -\frac{3r}{\sqrt{r^2+1}}\, ,\qquad
C_{tr\phi } = \frac{3r}{\sqrt{r^2+1}}\, ,\nonumber\\
C_{r\phi t} &= -\frac{6r}{\sqrt{r^2+1}}\, ,\qquad
C_{r\phi \phi } = 18r\,\Big( 1-\frac{1}{\sqrt{r^2+1}}\Big) \, ,
\end{align}
so that the Cotton scalars \eqref{Psi} are simply
\begin{align}
\Psi _0 &= \frac{\mu ^3}{6}\, ,\nonumber\\[1mm]
\Psi _1 &= 0\, ,\nonumber\\
\Psi _2 &= \frac{\mu ^3}{18}\, , \label{Goedel}\\[1mm]
\Psi _3 &= 0\, ,\nonumber\\
\Psi _4 &= -\frac{\mu ^3}{6}\, . \nonumber
\end{align}
It is easy to evaluate the invariants \eqref{Invariants_I_J} and \eqref{Invariants_G_H_N},
\begin{equation}
I = -\frac{\mu ^6}{27}\, ,\qquad
J = -\frac{2\mu ^9}{27^2}\, ,\qquad
G = 0\, ,\qquad
N = 0\, .
\end{equation}
Using the flow diagram in Fig.~\ref{Flow_diagram}, we obtain that the G\"odel spacetime
\eqref{Godel} is of algebraic \emph{type~D}, in agreement with the results of \cite{GHHM}.
By solving \eqref{Class_eq} with the Cotton scalars \eqref{Goedel} we obtain the complex roots
${K=\pm {\rm i}}$. Therefore, the two double CANDs are \emph{complex}, and the spacetime is actually of a subtype~D$_{\rm c}$. In fact, using the \emph{complex} null basis
\begin{align}\label{doubleCAND}
\boldk &=  \frac{\mu }{3\sqrt{2}\,r}\, \Big[ 2\, \complex \, (1-\sqrt{r^2+1})\, \partial _t
+r\sqrt{r^2+1}\, \partial _r -\complex \, \partial _\phi \Big] \, ,\nonumber \\
\boldl &= \frac{\mu }{3\sqrt{2}\,r}\, \Big[ 2\,\complex \, (1-\sqrt{r^2+1})\, \partial _t
-r\sqrt{r^2+1}\, \partial _r -\complex \, \partial _\phi \Big] \, ,\\
\boldm &= \frac{\mu }{3} \,\, \complex \, \partial _t \, ,\nonumber
\end{align}
the \emph{real} Cotton scalars take the canonical form
\begin{equation}
\Psi _0 = 0\, ,\qquad \Psi _1 = 0\, ,\qquad \Psi _2 = -\frac{\mu ^3}{9}\, ,\qquad \Psi _3 = 0\, ,\qquad \Psi _4 = 0 \, ,
\end{equation}
with only $\Psi_2$ non-vanishing. Therefore, \emph{both} $\boldk$ and $\boldl$ given by \eqref{doubleCAND} are \emph{double CANDs}.

\subsubsection{Type~N spacetime}

An example of type~N metric was given by Eqs.~(128)--(130) in \cite{GHHM}, namely the orthonormal basis
\begin{align}
\boldsymbol{\omega }^0 &= \e ^{\mu y/2}\big[ (1+\tfrac{1}{2}\e ^{-\mu y})\, \dd t +(1-\tfrac{1}{2}\e ^{-\mu y})\, \dd x \big] \, ,\nonumber \\
\boldsymbol{\omega }^1 &= \tfrac{1}{2}\e ^{-\mu y/2}(\dd t -\dd x)\, ,\\
\boldsymbol{\omega }^2 &= \dd y\, ,\nonumber
\end{align}
which yields the line element
\begin{equation}\label{typeN-metric}
\dd s^2 = -(1+\e ^{\mu y})\, \dd t^2 -2\e ^{\mu y}\, \dd t\, \dd x +(1-\e ^{\mu y})\, \dd x^2 +\dd y^2 \, .
\end{equation}
The corresponding null triad reads
\begin{align}\label{typeN-triad}
\boldk &= \tfrac{1}{\sqrt{2}}\,\e ^{\mu y/2}(\partial _t -\partial _x)\, ,\nonumber \\
\boldl &= \tfrac{1}{\sqrt{2}}\,\e ^{-\mu y/2}\Big[ (\e ^{\mu y}-1)\, \partial _t -(\e ^{\mu y}+1)\, \partial _x \Big] \, ,\\
\boldm &= \partial _y \, .\nonumber
\end{align}
The only non-vanishing components of the Cotton tensor are
\begin{equation}
C_{tyt} = C_{tyx} = C_{xyt} = C_{xyx} = -\tfrac{1}{2}\mu ^3 \e ^{\mu y} \, ,
\end{equation}
which projected onto the triad \eqref{typeN-triad} give
\begin{equation} \label{TypeN_Psi4}
\Psi _0 = \Psi _1 = \Psi _2 = \Psi _3 = 0 \, ,\qquad \Psi _4 = -\mu ^3\, .
\end{equation}
In view of the definition of invariants \eqref{Invariants_I_J} and \eqref{Invariants_G_H_N} we obtain
\begin{equation} \label{TypeN_Invariants}
I = 0 =J \, , \qquad G = 0 = H\, ,
\end{equation}
and using the flow  diagram in Fig.~\ref{Flow_diagram} we immediately see that the spacetime \eqref{typeN-metric} is of type~N, in agreement with \cite{GHHM}. Moreover, it is clear that the Cotton scalars \eqref{TypeN_Psi4} are already in the canonical form with only the scalar $\Psi _4$ non-trival, see Table~\ref{Tab:algebraic-types}. The vector $\boldk$ given by \eqref{typeN-triad} is thus a \emph{quadruple CAND}.

Notice that by performing transformations ${u=t+x}$ and ${r=\tfrac{1}{2}(t-x)}$, the metric \eqref{typeN-metric} takes the canonical Brinkmann form of pp-waves \cite{Podolsky&Papajcik, PodolskySvarcMaeda:2019}
\begin{equation}\label{typeN-metric-Brinkmann}
\dd s^2 =  \dd y^2 -2\, \dd u\, \dd r + a\, \dd u^2\, ,
\end{equation}
with ${\boldk \propto\,\partial _r}$  and specific metric function ${a=-\e ^{\mu y}}$ which depends only on the transverse spatial coordinate~$y$. Actually, it is a VSI spacetime with pure radiation (the only non-trivial component of the energy-momentum tensor is ${T_{uu}=\frac{\mu^2}{16\pi}\,\e ^{\mu y}}$), see Eqs.~(101), (102) in \cite{PodolskySvarcMaeda:2019}.

\subsubsection{Type~O spacetime}

Finally, we consider spherically symmetric solution with perfect fluid of a constant density $\rho$ and pressure $p(r)$,
\begin{equation}
\dd s^2 = -N^2\, \dd t^2 +\frac{\dd r^2}{F^2} +r^2\, \dd \phi ^2 \, .
\end{equation}
The metric functions are
\begin{align}
N(r) &= \frac{c_1}{\rho + p(r)}\, , \nonumber\\
F^2(r) &= c_2 -(\ell\, \rho +\Lambda )\,r^2\, ,\\
p(r) &= \frac{c_3\, (\ell\,\rho +\Lambda )\, F(r)+c_3^2\, \ell\, \Lambda +\rho\,F^2(r)}{c_3^2\, \ell ^2-F^2(r)} \, , \nonumber
\end{align}
where $c_1, c_2, c_3, \ell$ are constants, see Eqs.~(134), (139), (140) and (142) in \cite{GHHM}. The Cotton tensor for this metric identically vanishes (${C_{abc} = 0}$), so that its projections onto a null triad give
\begin{equation}
\Psi_{\rm A}  = 0\quad\hbox{for all A}\,.
\end{equation}
The spacetime is \emph{conformally flat}, that is of \emph{algebraic type~O}.

\section{Summary}

We introduced a useful approach to algebraic classification of 2+1 geometries, assuming no particular field equations. It is based on projecting the Cotton tensor onto a null triad. The corresponding five real Cotton scalars~$\Psi_{\rm A}$ (which are the 2+1 analogue of well-known 3+1 Newman--Penrose curvature scalars constructed from the Weyl tensor) then simply determine the algebraic types I, II, III, N, D and O by their gradual vanishing, starting with those of the highest boost weight, see Table~\ref{Tab-classification}. Moreover, such a classification is directly related to the specific multiplicity of the Cotton-aligned null directions (CANDs) and to the Bel--Debever criteria, see Table~\ref{Tab:algebraic-types} and Table~\ref{Tab:DB-criteria}, respectively. We also derived a  synoptic algorithm of the algebraic classification based on the polynomial curvature invariants (\ref{Invariants_I_J}), (\ref{Invariants_G_H_N}), see Fig.~\ref{Flow_diagram} (or Table~\ref{Tab:Algorithm-special-case} when ${\Psi_4 =0 =\Psi _0}$).

Using the bivector decomposition, we showed that our method is equivalent to the previously introduced Petrov-type classification of 2+1 spacetimes based on the eigenvalue problem and respective canonical Jordan form of the Cotton--York tensor, see Table~\ref{Tab:Petrov-Types} and Table~\ref{Tab:Normal-Types}.

In addition, we introduced a refinement of the algebraic types into the subtypes~I$_{\rm r}$, II$_{\rm r}$, D$_{\rm r}$ (for which all CANDs are real) and subtypes~I$_{\rm c}$, II$_{\rm c}$, D$_{\rm c}$ (for which some of the CANDs are complex). The subtype~I$_{\rm c}$ is related to the Class~I' defined in \cite{GHHM, Garcia:2017}, and the subtypes I$_{\rm r}$ and I$_{\rm c}$ correspond to the Petrov--Segre types I$_{\mathbb R}$ and I$_{\mathbb C}$ in TMG defined in \cite{ChowPopeSezgin:2010a}.

In final Sec.~\ref{sec:example} we demonstrated the practical usefulness of our novel method on several explicit examples of various algebraic types. We hope that it will prove to be helpful for classification and analysis of other interesting spacetimes in 2+1 gravity.

\section*{Acknowledgments}

This work was supported by the Czech Science Foundation Grant No.~GA\v{C}R 23-05914S.

\end{document}